\RequirePackage{etoolbox}
\csdef{input@path}%
{%
 {sty/},
 {img/},
}

\documentclass{elsevierbook}

\usepackage{graphicx}

\usepackage[normalem]{ulem}

\newcommand{\appropto}{\mathrel{\vcenter{
  \offinterlineskip\halign{\hfil$##$\cr
    \propto\cr\noalign{\kern2pt}\sim\cr\noalign{\kern-2pt}}}}}

\begin{document}

\Frontmatter

\Mainmatter

\begin{frontmatter}

\chapter{Kinks in buckled graphene uncompressed and compressed in the longitudinal direction}


\begin{aug}
\author[addressrefs={ad1,ad2}]%
{%
\fnm{Ruslan D.} \snm{Yamaletdinov}%
\footnote{Email: yamaletdinov@niic.nsc.ru}%
}%
\author[addressrefs={ad3}]%
{%
\fnm{Yuriy V.} \snm{Pershin}%
\footnote{Email: pershin@physics.sc.edu}%
}%
\address[id=ad1]%
{%
Nikolaev Institute of Inorganic Chemistry SB RAS,
Novosibirsk, 630090, Russia
}%
\address[id=ad2]%
{%
Boreskov Institute of Catalysis SB RAS,
Novosibirsk, 630090 Russia
}%
\address[id=ad3]%
{%
University of South Carolina,
Department of Physics and Astronomy,
Columbia, SC 29208, USA
}%
\end{aug}

\begin{abstract}
 In this Chapter we provide a review of the main results obtained in the modeling of graphene kinks and antikinks, which are elementary topological excitations of buckled graphene membranes. We introduce the classification of kinks, as well as discuss kink-antikink scattering, and radiation-kink interaction. We also report some new findings including {\it i}) the evidence that the kinetic energy of graphene kinks is described by a relativistic expression, and {\it ii}) demonstration of damped dynamics of kinks in membranes compressed in the longitudinal direction. Special attention is paid to highlight the similarities and differences between the graphene kinks and kinks in the classical scalar $\phi^4$ theory. The unique properties of graphene kinks discussed in this Chapter may find applications in nanoscale motion.
\end{abstract}

\begin{keywords}
\kwd{Buckled graphene}
\kwd{kink solution}
\kwd{$\phi^4$ model}
\end{keywords}

\end{frontmatter}

\section{Introduction}

Graphene kinks and antikinks are topological states of buckled graphene membranes introduced by us in Ref.~\cite{Yamaletdinov17b}. Up to the moment, only the studies of kink-antikink scattering~\cite{Yamaletdinov17b} and radiation-kink interaction~\cite{Yamaletdinov19a} have been reported.  However, these two publications already suggest a rich physics of topological excitations in buckled graphene and similar structures. The remarkable characteristics of graphene kinks and antikinks – the nanoscale size, stability, high propagation speed, and low energy dissipation – make them well positioned for applications in nanoscale motion.

Although graphene kinks are reminiscent the ones in the classical scalar $\phi^4$ theory~\cite{weinberg2012classical}, they are considerably more complex and require different approaches for their understanding. The complexity stems form the fact that the graphene membranes are two-dimensional objects described by multiple degrees of freedom. Importantly, even when the membrane motion is confined to one dimension, the transverse degrees of freedom can not be  neglected. The transverse degrees of freedom ``dress'' the longitudinal kinks in different ways leading to several types of graphene kinks compared to the single kink type in the classical scalar $\phi^4$ theory.

\begin{figure}[t]%
\begin{center}
\begin{tabular}{rc|c}
 &(a)&(b)\\
  $(\alpha,+)$-kink &\includegraphics[width=65mm]{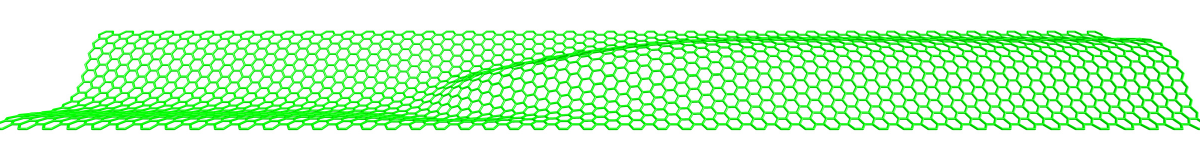}& \includegraphics[width=10mm]{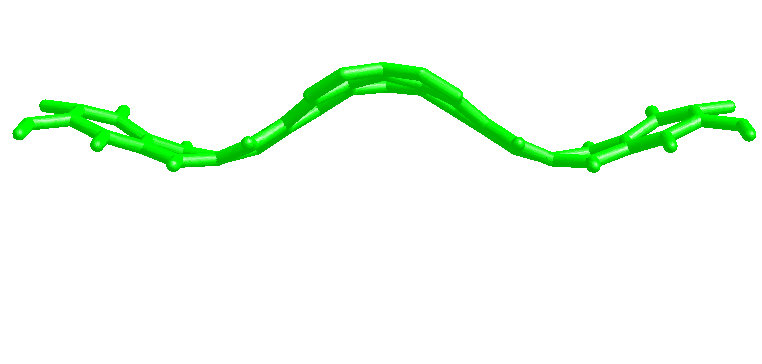}\\
  $(\alpha,-)$-kink &\includegraphics[width=65mm]{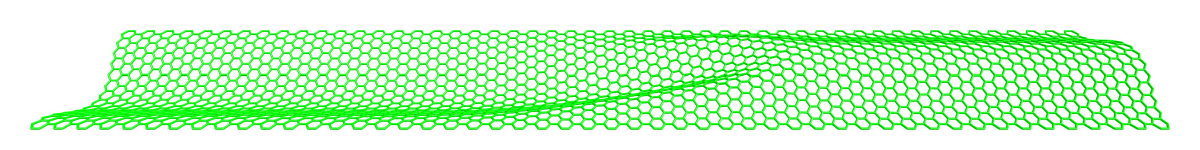}& \includegraphics[width=10mm]{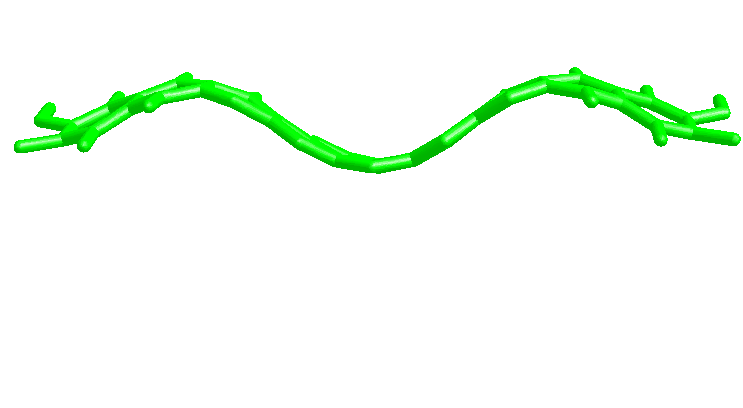} \\
  $(\beta,+)$-kink &\includegraphics[width=65mm]{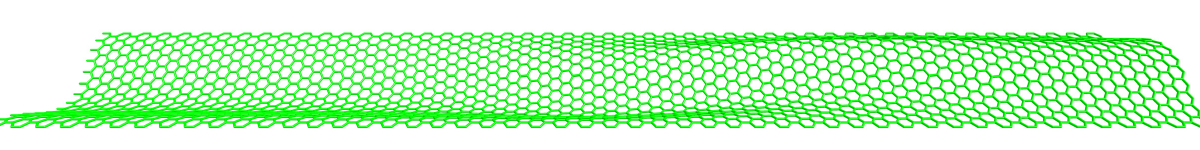} & \includegraphics[width=10mm]{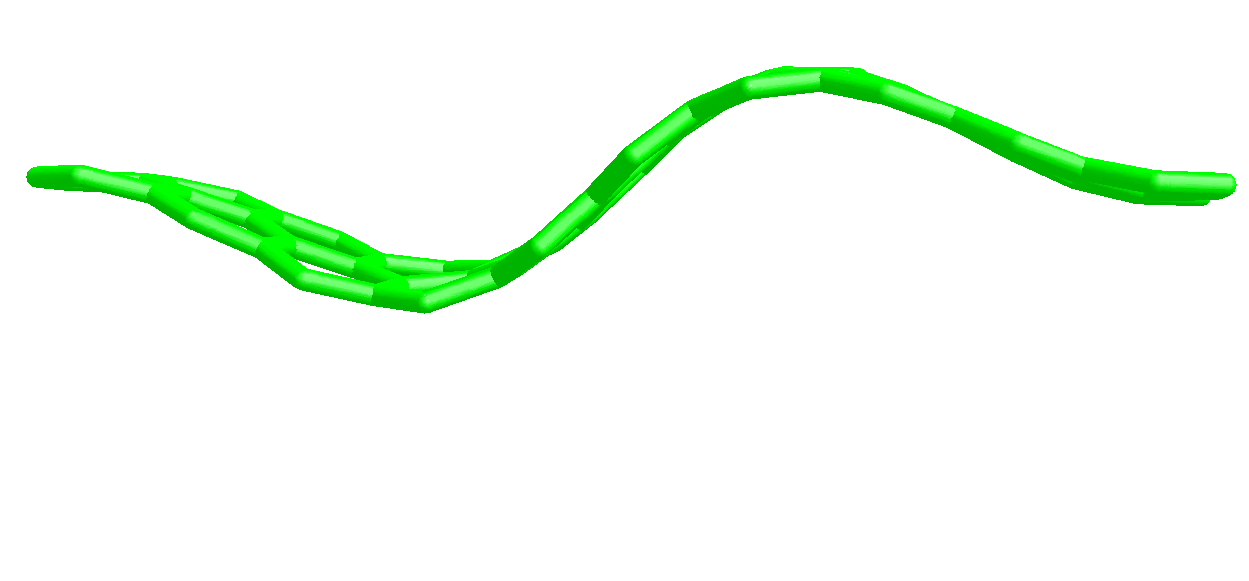} \\
  $(\beta,-)$-kink &\includegraphics[width=65mm]{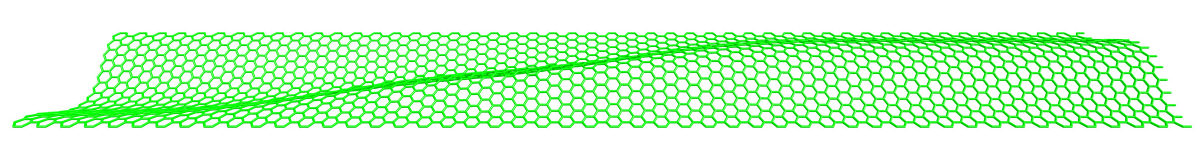}&  \includegraphics[width=10mm]{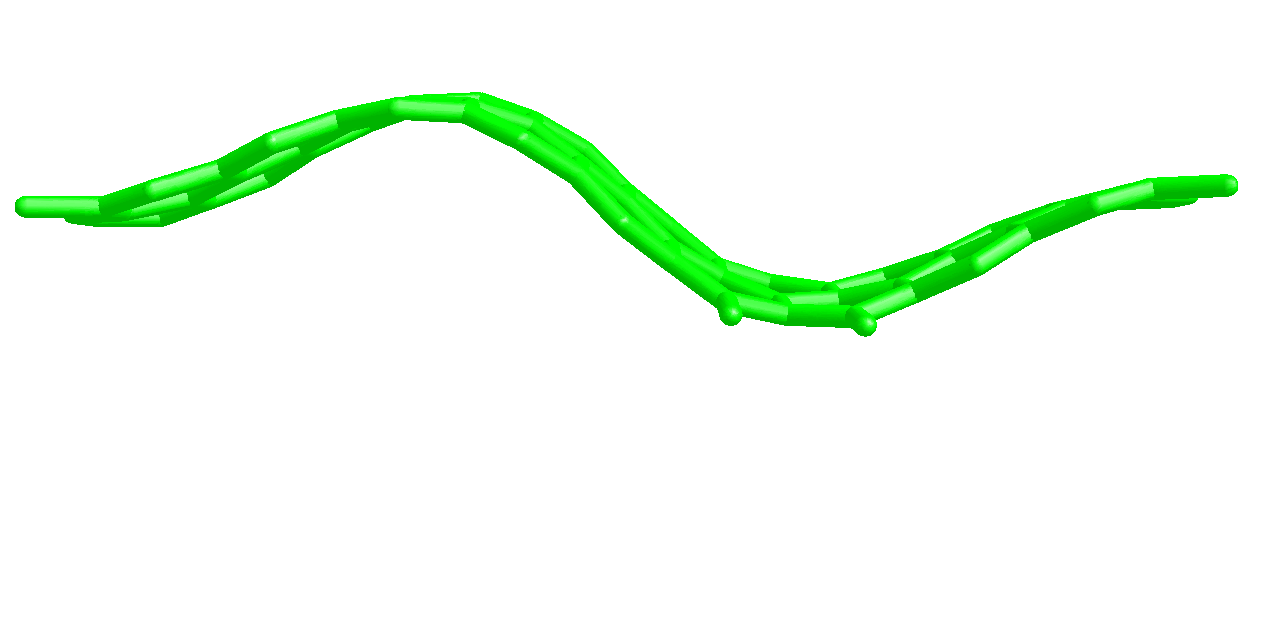} \\
  $(\alpha,+)$-antikink &\includegraphics[width=65mm]{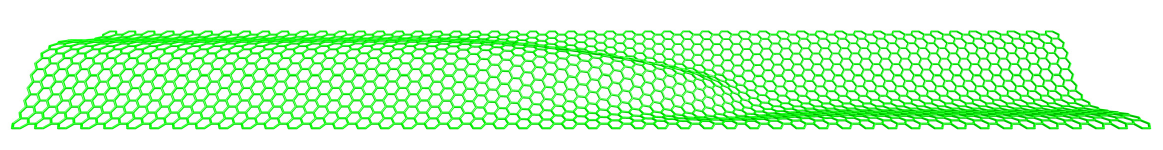}& \includegraphics[width=10mm]{fig1_1a}\\
  $(\alpha,-)$-antikink &\includegraphics[width=65mm]{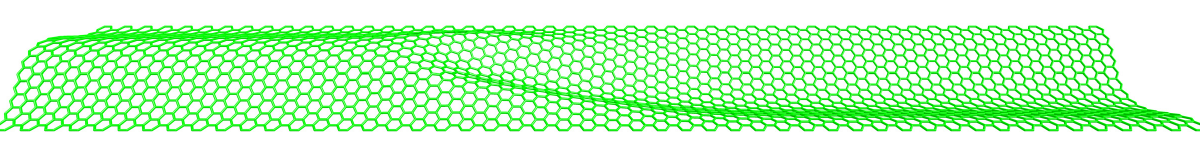}& \includegraphics[width=10mm]{fig1_2a} \\
  $(\beta,+)$-antikink &\includegraphics[width=65mm]{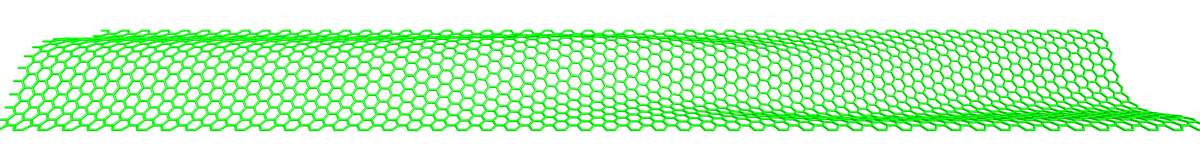}& \includegraphics[width=10mm]{fig1_3a} \\
  $(\beta,-)$-antikink &\includegraphics[width=65mm]{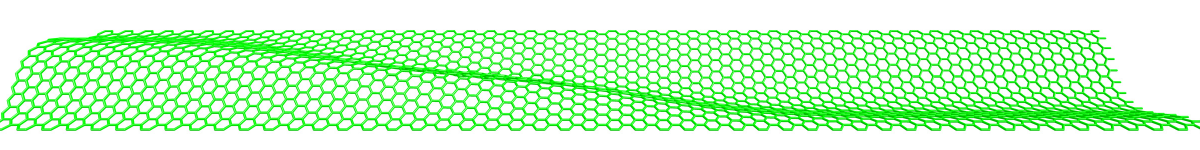}& \includegraphics[width=10mm]{fig1_4a} \\
     &\includegraphics[width=23mm]{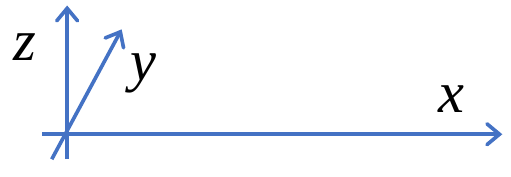}& \includegraphics[width=13mm]{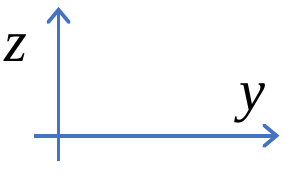} \\
\end{tabular}
\end{center}
\caption{Kinks and antikinks of buckled graphene (a), and their transverse cross-sections (b).}
\label{fig:1}
\end{figure}

To be more specific, consider a buckled graphene membrane (nanoribbon) over a trench assuming that the trench length $L$ (in the $x$-direction) is much longer than its width $d$ (in the $y$-direction). The buckled graphene has two stable configurations minimizing its energy: the uniform buckled up and buckled down states. Graphene kinks and antikinks are excited states of membrane connecting these uniform states, see Fig.~\ref{fig:1}. Fundamentally, kinks and antikinks differ by topological charge~\cite{vachaspati2006kinks,weinberg2012classical}
\begin{equation}
  Q_{\textnormal{top}} = \frac{1}{2z_0}\left(z(x \gg x_0,0)-z(x\ll x_0,0) \right) \label{eq:Qtop}
\end{equation}
such that $ Q_{\textnormal{top}}=1$ for kinks and $ Q_{\textnormal{top}}=-1$ for antikinks. In Eq.~(\ref{eq:Qtop}), $x_0$ is the position of kink or antikink, $z(x,y)$ is membrane deflection from the ($x,y$)-plane, and $z_0=|z(x\gg x_0,0)|=|z(x\ll x_0,0)|$ is the  deflection in the uniform buckled up or down state. The topological charge is defined by the boundary conditions (buckling directions very far left and very far right from the kink or antikink in Fig.~\ref{fig:1}), and does not depend on the transverse cross-section of membrane at any point. The states connecting the uniform buckled down membrane at $x\ll x_0$ and buckled up membrane at $x\gg x_0$  are kinks. The opposite states are antikinks.

Using molecular dynamics simulations, we have identified four types of kinks and four types of antikinks, see Fig.~\ref{fig:1}. Let us assign $\alpha(\beta)$-type to the kinks and antikinks with the symmetric (asymmetric) transverse cross section (in $y-z$ plane), and use $+(-)$ for kinks and antikinks with unturned (under the transformation $z\rightarrow -z$) transverse cross section (more details are given below). We note that all $\alpha$-type kinks and antikinks can be obtained from a single one via symmetry transformations. For instance, the transformation $z\rightarrow -z$ transforms $(\alpha,+)$-kink into  $(\alpha,-)$-antikink and vice-versa,  the transformation $x\rightarrow -x$ transforms $(\alpha,+)$-kink into  $(\alpha,+)$-antikink, etc. The same is true for $\beta$-type kinks and antikinks. We emphasize that the longitudinal cross-section (in $x-z$ plane at $y=0$) defines whether the state is a kink or antikink, while the transverse cross-section (in $y-z$ plane) defines its type $\alpha(\beta)$, $+(-)$. Additional details are provided in Sec.~\ref{sec:3a}.

Our previous work on graphene kinks~\cite{Yamaletdinov17b,Yamaletdinov19a} was based on molecular-dynamics simulations, which is a versatile tool to investigate mechanical properties of nanostructures. The success and easiness to use of molecular dynamics are related to the atomistic approach to describe the system of interest, availability of potentials that realistically describe interactions between atoms, and modest requirement to computational resources (compared to the density functional theory calculations). During the last decade, the systems comprising millions on atoms have been routinely modeled on modest computer clusters. Some billion-atom molecular dynamics simulations have been reported recently~\cite{shibuta2017heterogeneity,Jaewoon19a}.  We note that the elasticity theory (more specifically, the classical non-linear theories for plates~\cite{timoshenko1959theory} with strain limitations) offers an alternative  framework to the description of graphene elasticity~\cite{Samadikhah2007,Jiang2009}. An interesting future problem is to describe analytically the shape of graphene kinks, and find analytical expression for kink energy.

In this Chapter we review our previous results on graphene kinks~\cite{Yamaletdinov17b,Yamaletdinov19a} and extend them in two directions. First, we explore how the energy of graphene kinks relates to the membrane width, degree of buckling, and kink velocity. Second, we investigate the effect of  longitudinal stress on the dynamics of graphene kinks.

The Chapter is organized as follows. Section~\ref{sec:2} provides some preliminary information including $\it i$) the classical scalar $\phi^4$ model, which sets the framework for the discussion, and  $\it ii$) details of molecular dynamics simulations. Next, in Sec.~\ref{sec:3} we consider graphene kinks in longitudinally uncompressed graphene. Here, we introduce the nomenclature of graphene kinks, report our new results on kink energetics, as well as briefly review some of our past findings~\cite{Yamaletdinov17b,Yamaletdinov19a}. Section~\ref{sec:4} reports few selected results on graphene kinks in longitudinally compressed membranes. Finally, the conclusions and outlook are presented in Sec.~\ref{sec:5}.

\section{Preliminaries} \label{sec:2}

\subsection{$\phi^4$ model} \label{sec:2a}

To set the stage for a discussion of graphene kinks, let us briefly review kinks in the $\phi^4$ classical scalar field theory~\cite{weinberg2012classical,vachaspati2006kinks,kevrekidis2019dynamical}. For the last several decades this model has been widely used in several diverse branches of physics including statistical mechanics, condensed matter, topological quantum field theory~\cite{manton2004topological}, and cosmology~\cite{vilenkin2000cosmic}. In the area of condensed matter physics, $\phi^4$-kinks have been used to describe domain walls in ferroeletrics~\cite{aubry1974dynamical,krumhansl1975dynamics,Saxena06a} and ferromagnets, proton transport in hydrogen-bonded chains~\cite{Kashimori82a,Laedke85a,Peyrard87a}, and charge-density waves~\cite{Rice76a,RICE1979368}. We note that in recent years, the focus of attention has shifted to higher-$\phi$ models, such as $\phi^6$~\cite{gani2014kink,marjaneh2017multi}, $\phi^8$~\cite{gani2015kink,belendryasova2019scattering}, $\phi^{10}$ and $\phi^{12}$~\cite{khare2014successive}. Their stables states, however, are quite different compared to the ones that we observe in graphene.

The $\phi^4$ model is based on the Lagrangian
\begin{equation}\label{eq:Lagrangian}
\mathcal{L}=\frac{1}{2}\int \textnormal{d}\tilde{x} \left[\left( \frac{\partial \phi}{\partial \tilde{t}}\right)^2-\left( \frac{\partial \phi}{\partial \tilde{x}}\right)^2
-\frac{1}{2}(1-\phi^2)^2\right],
\end{equation}
where $\phi(\tilde{x},\tilde{t})$ is a real scalar field, $\tilde{x}$ and $\tilde{t}$ are dimensionless spatial coordinate and time, respectively. We use tildes to distinguish dimensionless quantities from the dimensional ones; for an example of  $\phi^4$ model written in dimensional units, see Ref.~\cite{currie80}. Lagrangian (\ref{eq:Lagrangian}) leads to the Euler-Lagrange equation of motion of the form
\begin{equation}\label{eq:phi4}
  \frac{\partial^2 \phi}{\partial \tilde{t}^2}- \frac{\partial^2 \phi}{\partial \tilde{x}^2}= \phi-\phi^3.
\end{equation}
Eq.~(\ref{eq:phi4}) has two stable constant solutions $\phi=\pm 1$, and one unstable $\phi=0$.
The kink and antikink solutions connect the two stable ground state values, and, therefore, they are topologically stable.
Mathematically, these are given by
\begin{equation}\label{eq:Kink}
 \phi_{K(A)}=\pm \tanh \left( \frac{\tilde{x}-\tilde{V}\tilde{t}-\tilde{x}_0}{\sqrt{2(1-\tilde{V}^2)}}\right),
\end{equation}
where $\pm$ signs correspond to the kink and antikink, respectively, $\tilde{V}$ is the dimensionless velocity ($|\tilde{V}|<1$), and $\tilde{x}_0$ is the dimensionless position of the kink/antikink center at the initial moment of time $\tilde{t}=0$. According to Eq.~(\ref{eq:Kink}), $\phi^4$ kinks and antikinks move without dissipation at a constant velocity. Moreover, their kinetic energy is expressed by a relativistic formula~\cite{currie80}. In the past, the above model was applied to different physical situations, where the system can be effectively described by a 1D model with a double well potential. The physical meaning of $\phi(\tilde{x},\tilde{t})$ is application-specific.
For instance, in Ref.~\cite{Saxena06a}, $\phi$ was used to describe the magnetic order parameter, while in the present Chapter it is used to represent the membrane deflection.

\begin{figure}[t]%
 \includegraphics*[width=55mm]{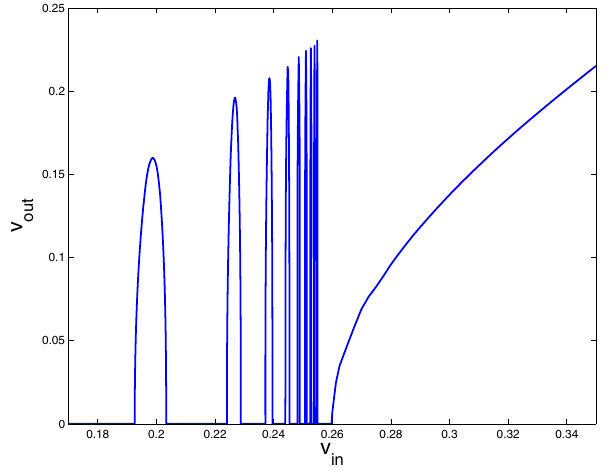}
\caption{Kink-antikink collision in $\phi^4$ model: Final velocity as a function of the initial kink velocity~\cite{Goodman05a}. Reprinted with permission from Ref.~\cite{Goodman05a}.
}
\label{fig:2}
\end{figure}

A very famous result in $\phi^4$ theory is the resonant structure of the kink-antikink scattering~\cite{CAMPBELL19831}. Numerical simulations have shown that intervals of initial velocity for which the kink and antikink capture one another alternate with regions for which kink and antikink separate infinitely~\cite{CAMPBELL19831}, see Fig.~\ref{fig:2}. The main feature of the latter regions are two- and higher-bounce resonances in which kink and antikink form a quasi-stable state for some short period of time. Within this process, the kinetic energy is first transferred into internal oscillation mode(s), and realized at a later time (such as on the second bounce). The early work on two-bounce resonance is attributed to Campbell and co-authors~\cite{CAMPBELL19831}. Higher-order resonances were identified by Anninos, Oliveira, and Matzner~\cite{Anninos91a}.

Below we demonstrate that the solutions (\ref{eq:Kink}) of Eq. (\ref{eq:phi4}) are relevant to graphene kinks and antikinks. Eq.~(\ref{eq:Kink}) can be considered as zero-order approximation to the deflection of the central line ($y=0$) of membrane atoms from ($x,y$)-plane. Specifically, the graphene kinks are qualitatively related to the $\phi^4$ kinks through the following correspondence: the $x$-coordinate (along the membrane) corresponds to $\tilde{x}$, the central line deflection $z(x,y=0,t)$ plays the role of $\phi(\tilde{x},\tilde{t})$, and the
 graphene kink velocity $V$ corresponds to $\tilde{V}$.
In Sec.~\ref{sec:3b} we show that the relativistic formula fits well the kinetic energy of graphene kinks. Moreover, preliminary data indicates the possibility of bounce resonances in graphene kink-antikink scattering (under certain conditions).


\subsection{Molecular dynamics simulations} \label{sec:MD}

Our results on graphene kinks were obtained using NAMD2~\cite{phillips05} -- a
highly scalable massively parallel classical MD code\footnote{NAMD was developed by the Theoretical and Computational Biophysics Group in the Beckman Institute for Advanced Science and Technology at the University of Illinois at Urbana-Champaign.} -- with optimized CHARMM-based force field~\cite{Best2012} for graphene atoms (see description below and Ref.~\cite{Yamaletdinov2018}). Some of our results were verified with a more comprehensive Tersoff potential~\cite{Tersoff1988} using LAMMPS code~\cite{plimpton1995fast}. This method has confirmed the validity of our simulations with NAMD2.

\begin{figure}[h]%
\includegraphics[width=65mm]{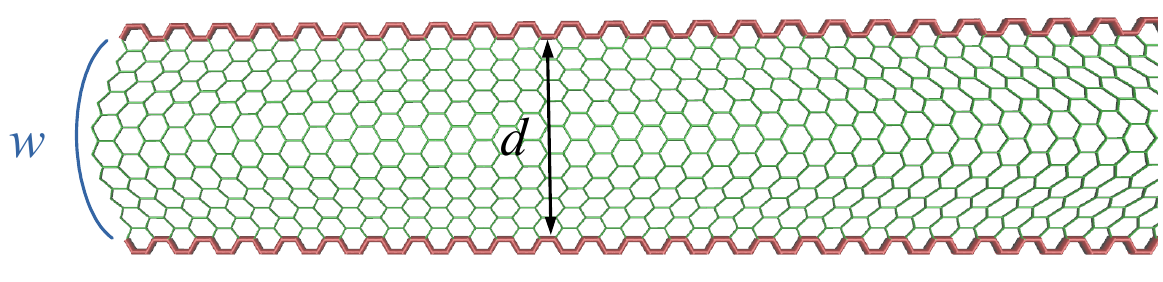}
\caption{Geometry of buckled graphene nanoribbon. The red armchair edges represent the location of  frozen atoms.}
\label{fig:geom}
\end{figure}

We simulated the dynamics of graphene nanoribbons (membranes) of a length $L$ and width $w$.
We used  clamped boundary conditions for the longer armchair
edges and free boundary conditions for the shorter edges, see Fig.~\ref{fig:geom}. To implement the clamped boundary conditions,
the two first lines of carbon atoms of longer edges were fixed.  Membranes were buckled by changing the
distance between the fixed sides from $w$ to $d < w$. In what follows for the sake of clarity we use ``ring'' units of width and length.
In our geometry one ring has the size (width $\times$ length) of $\sqrt{3}a_0 \times 3a_0 = 2.46$~\r{A} $\times 4.26$~\r{A}.

To simulate the graphene dynamics we used the optimized CHARMM-based force field~\cite{Best2012}, which includes 2-body spring bond, 3-body angular bond (including the Urey-Bradley term), 4-body torsion angle, and Lennard-Jones potential energy terms \cite{JCC21367}. All force-field constants have been optimized to match known properties of graphene. In particular, the AB stacking distance and energy of graphite \cite{Chen2013} have been used for choose the Lennard-Jones coefficients. The remaining parameters were optimized to reproduce the in-plane stiffness ($E_{2D}=342$ N/m), bending rigidity ($D=1.6$ eV) and equilibrium bond length ($a=1.421$ \r{A}) of graphene. All simulations were performed with 1 fs time step. The van der Waals interactions were gradually cut off starting at 10~\r{A} from the atom until reaching zero interaction 12~\r{A} away.

\vspace{0.1cm}

Several methods were used to explore the properties of graphene kinks:
\begin{itemize}
\item Method 1. To determine configurations and energies (Secs.~\ref{sec:3a} and \ref{sec:3b}), the Langevin dynamics of the initially flat compressed nanoribbon was simulated for 20
ps at $T = 293$ K using a Langevin damping parameter of
0.2 ps$^{-1}$ in the equations of motion. This simulation stage
was followed by 10000 steps of energy minimization. The temperature of 293 K was used in our initial simulations and resulted in very good results in terms of the configuration representativeness.

\item Method 2. To generate moving kinks (Secs.\ref{sec:3b} and \ref{sec:3c}), we applied a downward (in $-z$ direction) force to the atoms located at the distance up to about 3 rings from the shorter edge or edges. As the initial configuration, we used an optimized membrane in the uniform buckled up state. The dynamics was simulated for $20$~ps  at $T=0$~K without any temperature or energy control.
\item Method 3. To simulate the radiation-kink interaction (Sec.~\ref{sec:3d}), we applied a sinusoidal force to the atoms located at the distance up to about 3 rings from the shorter edge. As the initial configuration, we used an optimized membrane with a stationary kink located in the middle part of membrane. We performed a series of computations with $x-$, $y-$ or $z-$ directed force with amplitudes and vibration period in ranges $0-160$~pN/atom and $0-160$~fs, respectively.  The dynamics was simulated for $30$~ps at $T=0$~K without any temperature or energy control.
\end{itemize}

The interested reader can contact the authors directly for examples of input files of their molecular dynamics simulations.

\section{Kinks in longitudinally uncompressed graphene} \label{sec:3}

\subsection{Types of graphene kinks}  \label{sec:3a}

\begin{figure*}[t]%
 \includegraphics*[width=55mm]{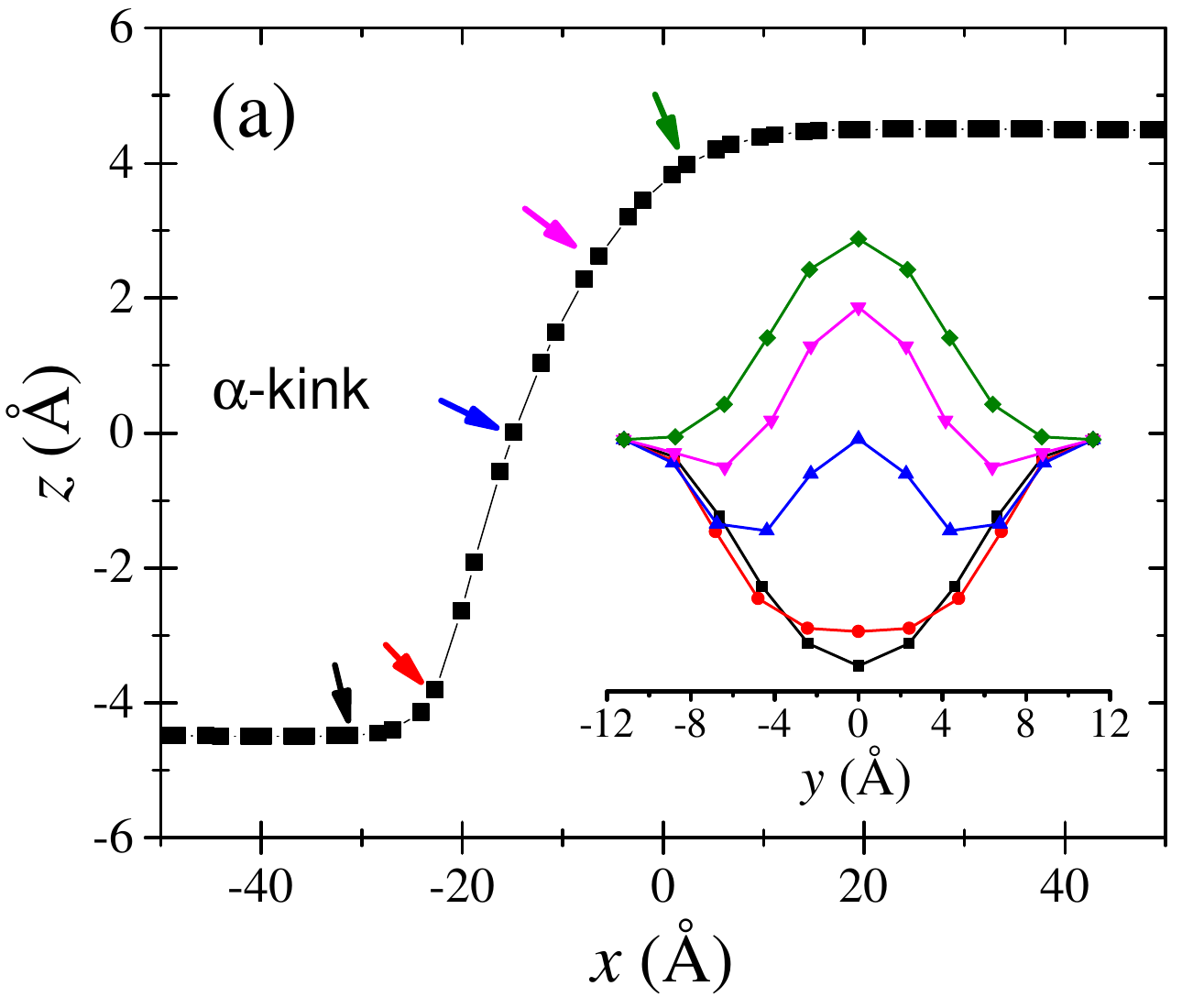}
 \includegraphics*[width=55mm]{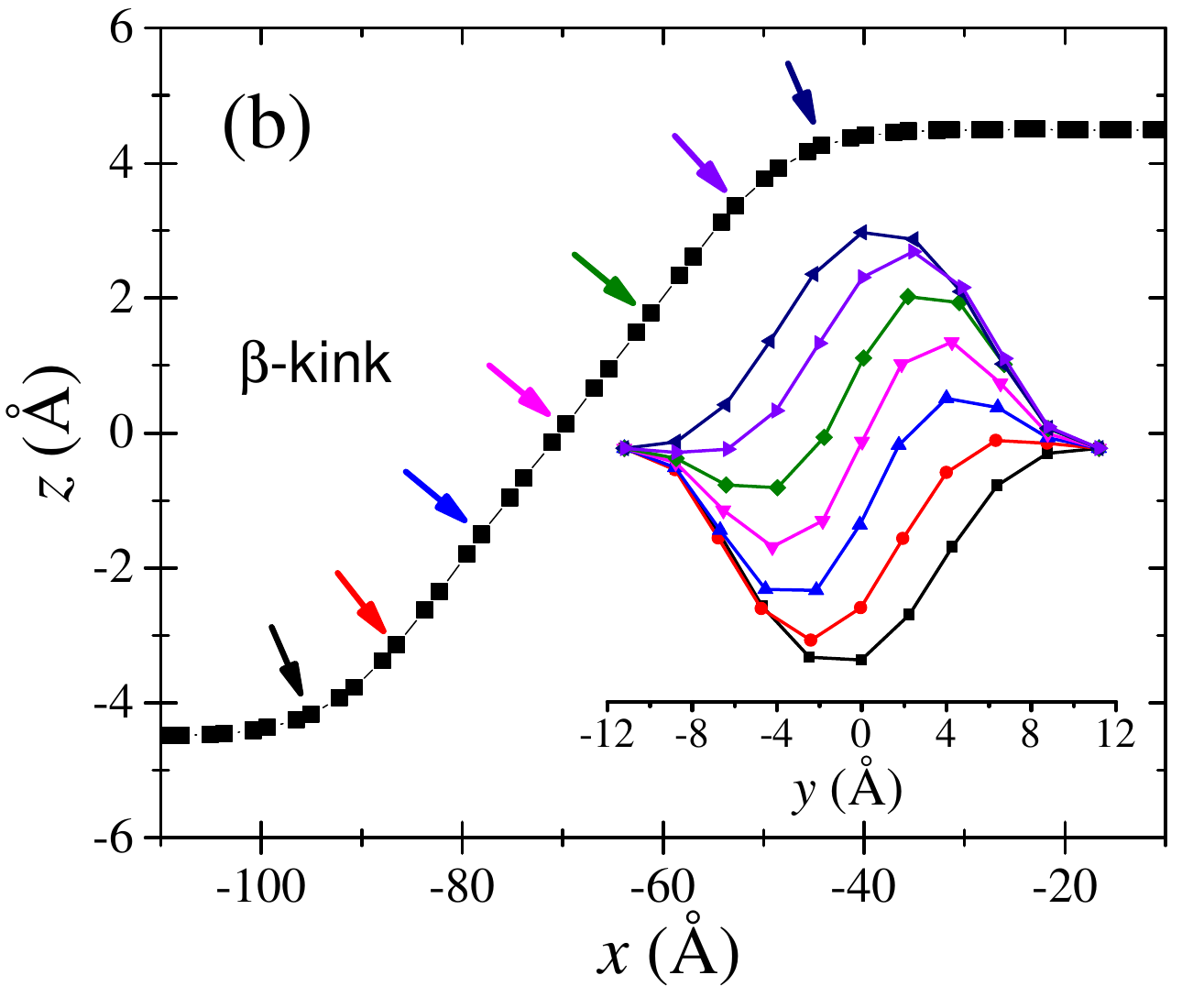}
\caption{(a) Symmetric and (b) non-symmetric kinks. The arrows show the positions of cross sections (insets).
These configurations were obtained using Method 1 in Sec.~\ref{sec:MD} for $d/w=0.9$, $w=9$ rings, and $L=100$ rings.}
\label{fig:3}
\end{figure*}

Our studies have revealed that there exists four types of graphene kinks and four types of antikinks, which are summarized in Fig.~\ref{fig:1}. In the {\it symmetric} kinks (referred as $\alpha$-kinks), the cross section in the transverse to the trench direction is symmetric (see Fig.~\ref{fig:3}(a)). Very approximately, close to the kink center $x_0$, the membrane deflection in symmetric kinks and antikinks as a function of $y$ can be represented by
\begin{equation}\label{eq:3a}
  z_\alpha(x\approx x_0,y)\appropto \pm \cos\left(\frac{3\pi y}{d} \right).
\end{equation}
The ``$+$'' and ``$-$'' $\alpha$-kinks correspond to $\pm$ in the above equation. According to this terminology, Fig.~\ref{fig:3}(a) represents an ($\alpha, +$)-kink.

Similarly, the {\it non-symmetric} kinks (referred as $\beta$-kinks) are characterized by a non-symmetric cross section. An example of non-symmetric kink is shown in Fig.~\ref{fig:3}(b). Again, very approximately, in the vicinity of the center  the cross section of $\beta$-kinks and antikinks is given by
\begin{equation}\label{eq:3b}
  z(x\approx x_0,y)\appropto \pm \sin\left(\frac{2\pi y}{d} \right).
\end{equation}
The ``$+$'' and ``$-$'' $\beta$-kinks correspond to $\pm$ in the above equation. The kink in Fig.~\ref{fig:3}(b) is thus a ($\beta, +$)-kink.

\begin{figure*}[bt]%
 \includegraphics*[width=55mm]{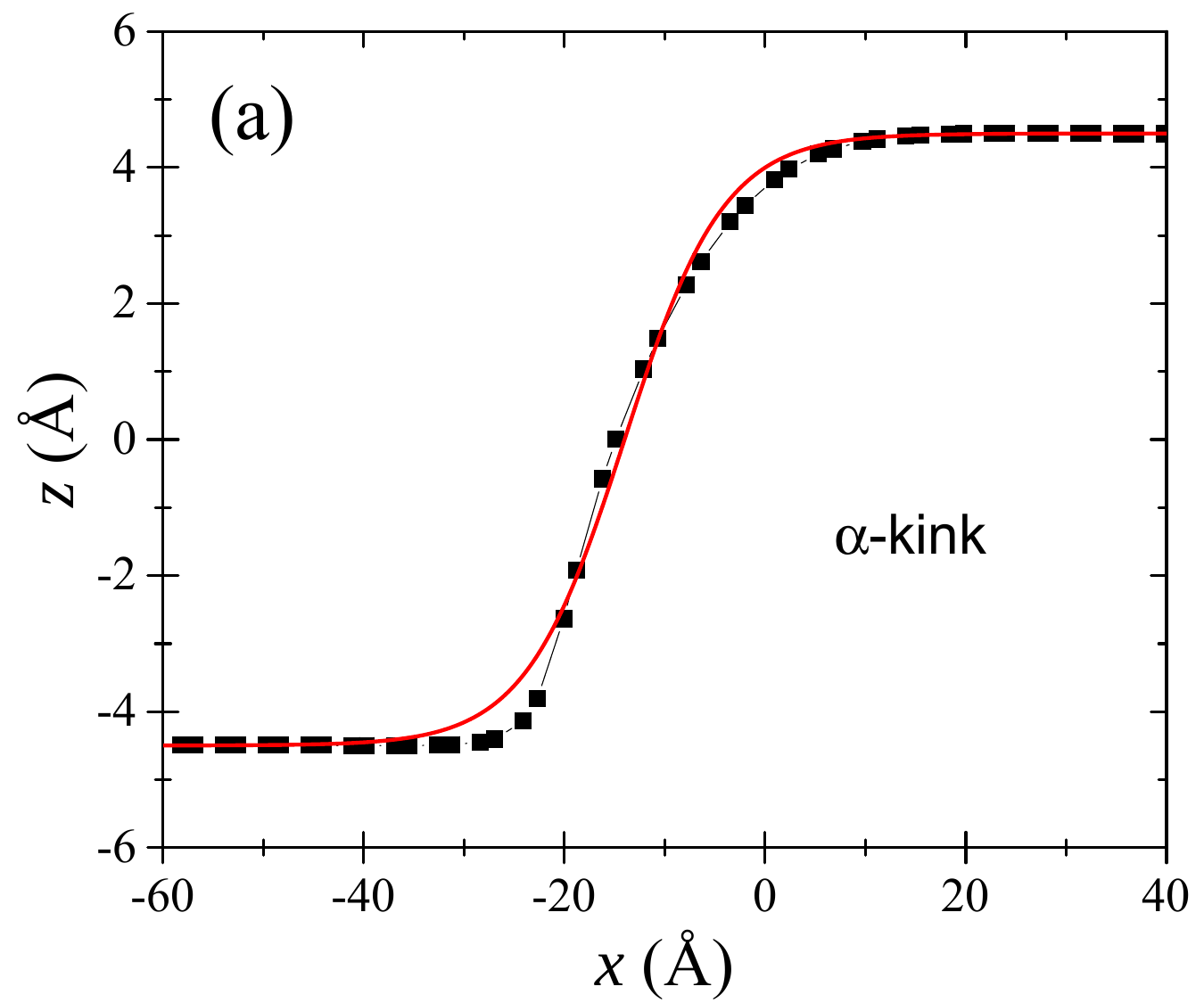}
 \includegraphics*[width=55mm]{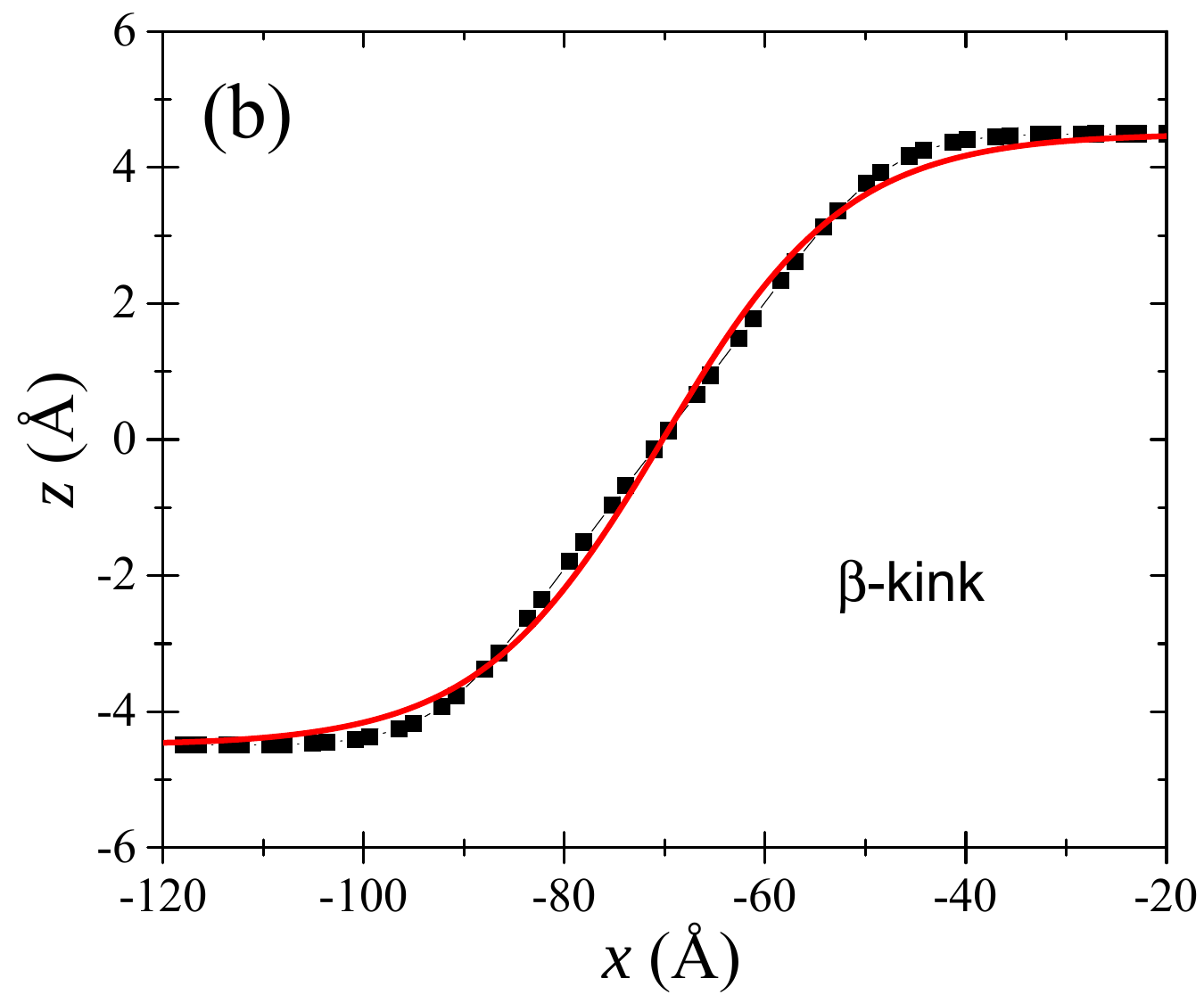}
\caption{Comparison of the graphene and $\phi^4$ kinks (red solid line).
To plot $\phi^4$ kinks we used Eq.~(\ref{eq:Kink}) with a scaled amplitude, $V=0$, and $x$ scaled by 7~\AA~in (a) and 13~\AA~in~(b). The graphene kinks are the same as in Fig.~\ref{fig:3}.
}
\label{fig:fit}
\end{figure*}

In Fig.~\ref{fig:fit}, the graphene and $\phi^4$ kinks (Eq.~(\ref{eq:Kink})) are superimposed for comparison. This plot indicates that the longitudinal cross sections of
$\alpha$- and $\beta$-kinks can not be ideally described by the ideal $\phi^4$ model. We also emphasize that
the $\beta$-kinks are approximately two times wider in $x$-direction compared to $\alpha$-kinks. Moreover, $\alpha$-kinks are slightly non-symmetric in the longitudinal direction.

\subsection{Kink energy} \label{sec:3b}

\subsubsection{Stationary kinks}

\begin{figure}[h]%
\includegraphics[width=55mm]{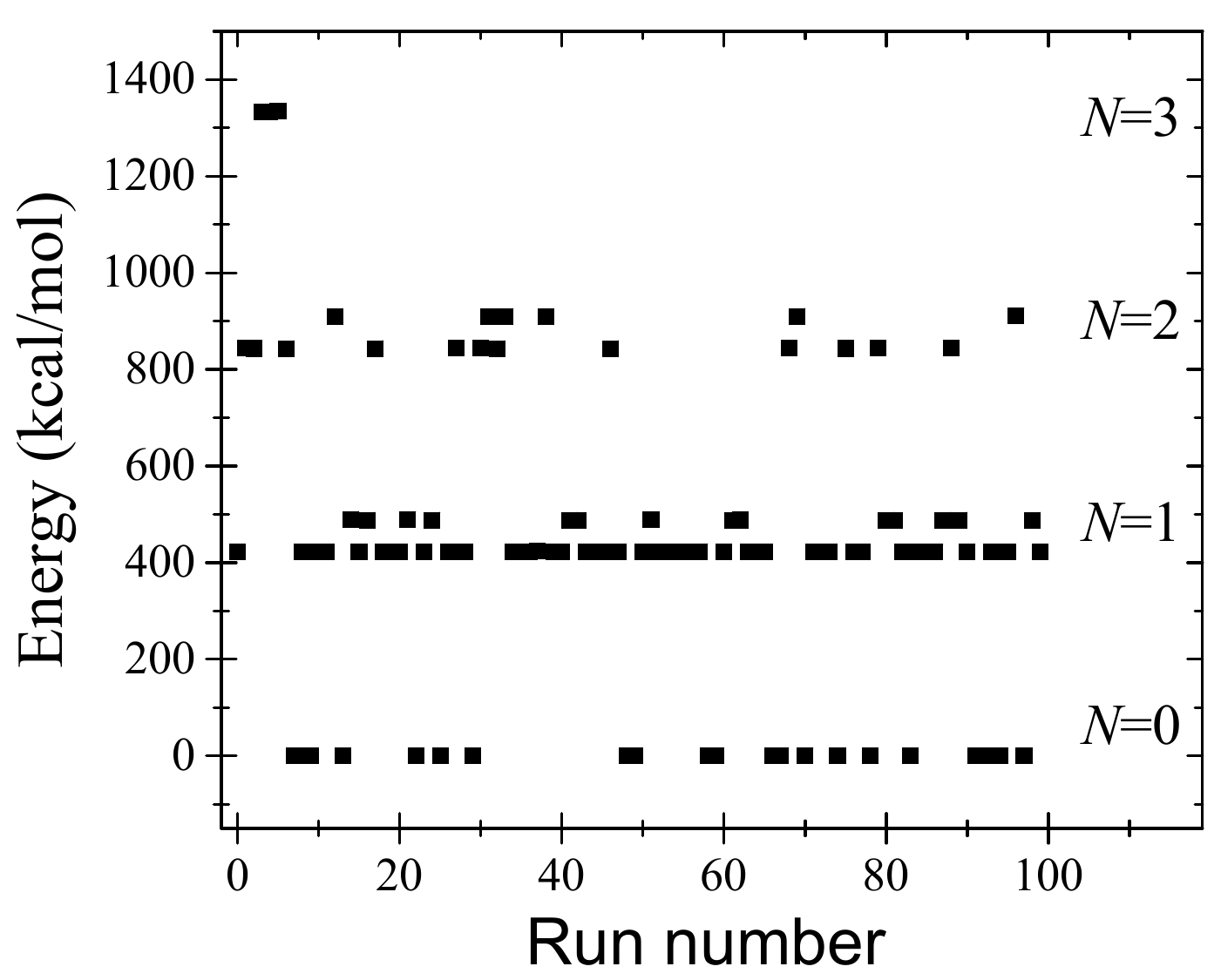}
\caption{Final energy, counted from the energy of the uniform buckled state, calculated using Method 1 in
100 independent runs.  Here, $N$ is the number of kinks
and antikinks in the final conformation for each run. This plot was obtained for a $w=13$ rings membrane. }
\label{fig:5}
\end{figure}

The energies of stationary  kinks were calculated using the dynamics/energy minimization approach for several values of $w$ and $d/w$ (Method 1 in Sec.~\ref{sec:MD}). The use of finite temperature dynamics in  Method 1 enables sampling the stable membrane conformations and their energies~\cite{yamaletdinov2017finding}. An example of our results is presented in Fig.~\ref{fig:5}. Here, the final conformation energies are plotted for 100 independent runs. Due to the stochastic nature of molecular dynamics simulations at finite temperature, the results are different in different runs.

Importantly, the final state energies in Fig.~\ref{fig:5}  are discrete. The lowest possible energy $N=0$ is the ground state energy in which the membrane is uniformly buckled up or down. The excited state energies correspond to the membrane with $N$ kinks and/or antikinks. While each kink/antikink contributes approximately the same amount of energy to the total energy, there is a difference in the energies of $\alpha$- and $\beta$-kinks. Due to this difference, the lines with the same $N$ are splitted. When kinks/antikinks are sufficiently far from the edges, and spaced apart enough to neglect their interaction, the total energy can be writen as
\begin{equation}\label{eq:Etotal}
  E_{tot}=E_0+(N_{K,\alpha}+N_{A,\alpha})E_{\alpha}+(N_{K,\beta}+N_{A,\beta})E_{\beta},
\end{equation}
where $E_0$ is the ground state energy, $E_{\alpha(\beta)}$ is the energy of $\alpha(\beta)$-kink at rest, $N_{K(A),\alpha}$ is the number of $\alpha$-kinks (antikinks), and $N_{K(A),\beta}$ is the number of $\beta$-kinks (antikinks), and $N=N_{K,\alpha}+N_{A,\alpha}+N_{K,\beta}+N_{A,\beta}$.
Clearly, the level splitting increases with $N$ as $N+1$. We note that in some calculations the minimization occasionally resulted in a kink or antikink trapped at the boundary with energy different than $E_{\alpha(\beta)}$. Such cases were rejected by manual inspection.

\begin{figure*}[t]%
 \includegraphics*[width=55mm]{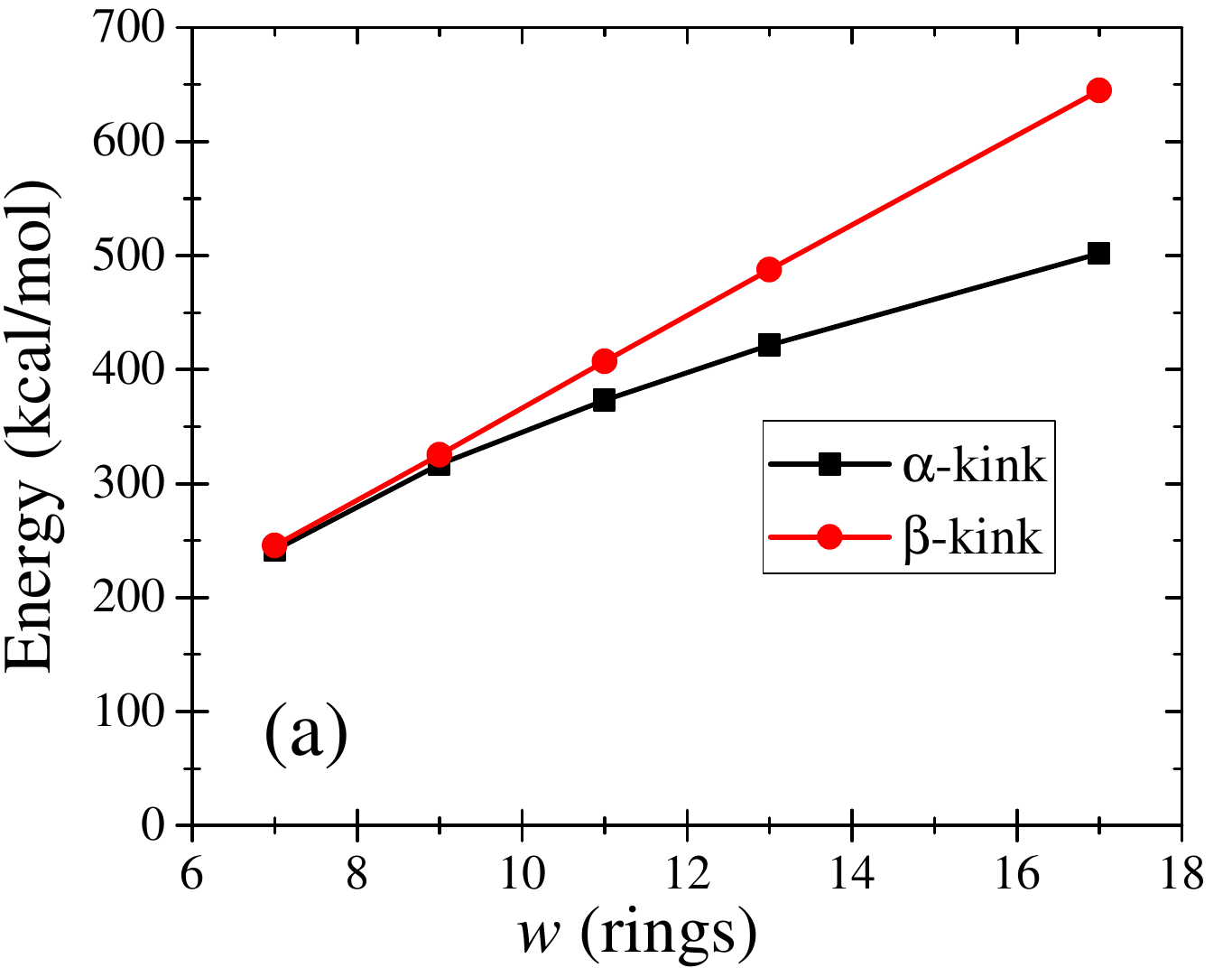}
 \includegraphics*[width=55mm]{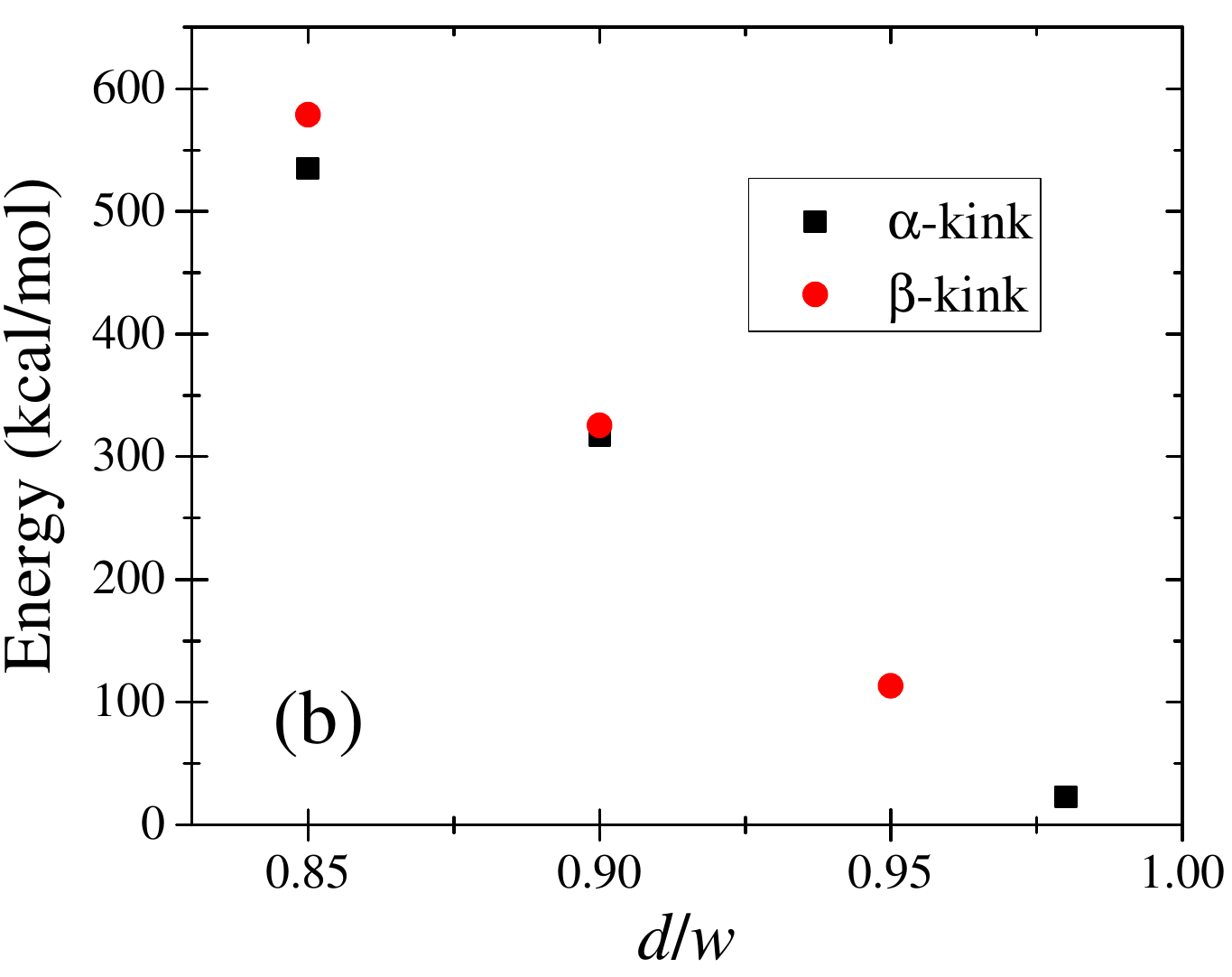}
\caption{Energies of symmetric and non-symmetric kinks plotted at fixed (a) $d/w=0.9$ and (b) $w=9$ rings.
In (b), we observed only non-symmetric kinks at $d/w=0.95$, and only symmetric ones at $d/w=0.98$ (in both cases, in 100 runs). }
\label{fig:4}
\end{figure*}

It is interesting to investigate how the kink energy changes with $w$ and $d/w$. For this purpose, we performed Method 1 simulations for selected values of $w$ and $d/w$ (similar to the ones reported in Fig.~\ref{fig:5}) and extracted kink energies from these simulations. Fig.~\ref{fig:4}(a) shows that at a fixed ratio of $w/d=0.9$ the energy of $\alpha$-kink is always smaller than the energy of $\beta$-kink. This plot demonstrates that the energy of $\beta$-kink increases almost linearly with the channel width, while the increase of $\alpha$-kink energy is non-linear and not so fast.

Fig.~\ref{fig:4}(b) demonstrates that at a fixed $w$, the kink energies decrease with increase of $d/w$. This is expected behavior. Unfortunately, the fine understanding of Fig.~\ref{fig:4} features is not possible based solely on molecular dynamics simulations. The combination of analytical and numerical approaches would be an ideal route to accomplish this goal.

\subsubsection{Moving kinks}

Another non-trivial task is to understand the properties of moving kinks. The classical scalar $\phi^4$ theory predicts that the kink kinetic energy
 is expressed by the relativistic formula~\cite{currie80}
\begin{equation}\label{eq:Kink_Ek}
 E_k=mC^2\left( \frac{1}{\sqrt{1-V^2/C^2}}-1\right),
\end{equation}
where $m$ is the effective kink mass, $C$ and $V$ are the characteristic and kink velocities, respectively.
In this subsection we demonstrate numerically that the relativistic expression (\ref{eq:Kink_Ek}) provides a much better fit to the kink kinetic energy compared to the classic expression. This strongly indicates that the  kinetic energy of the kink is described by the relativistic expression~\cite{currie80}. One of the major consequences is that the velocity of kink cannot exceed its ``speed of light'' $C$, which has no relation to the real speed of light, but enters in the same way in the kinetic energy of the kink as the real speed of light in the kinetic energy of particles moving at relativistic velocities.


\begin{figure*}[h]
\includegraphics*[width=55mm]{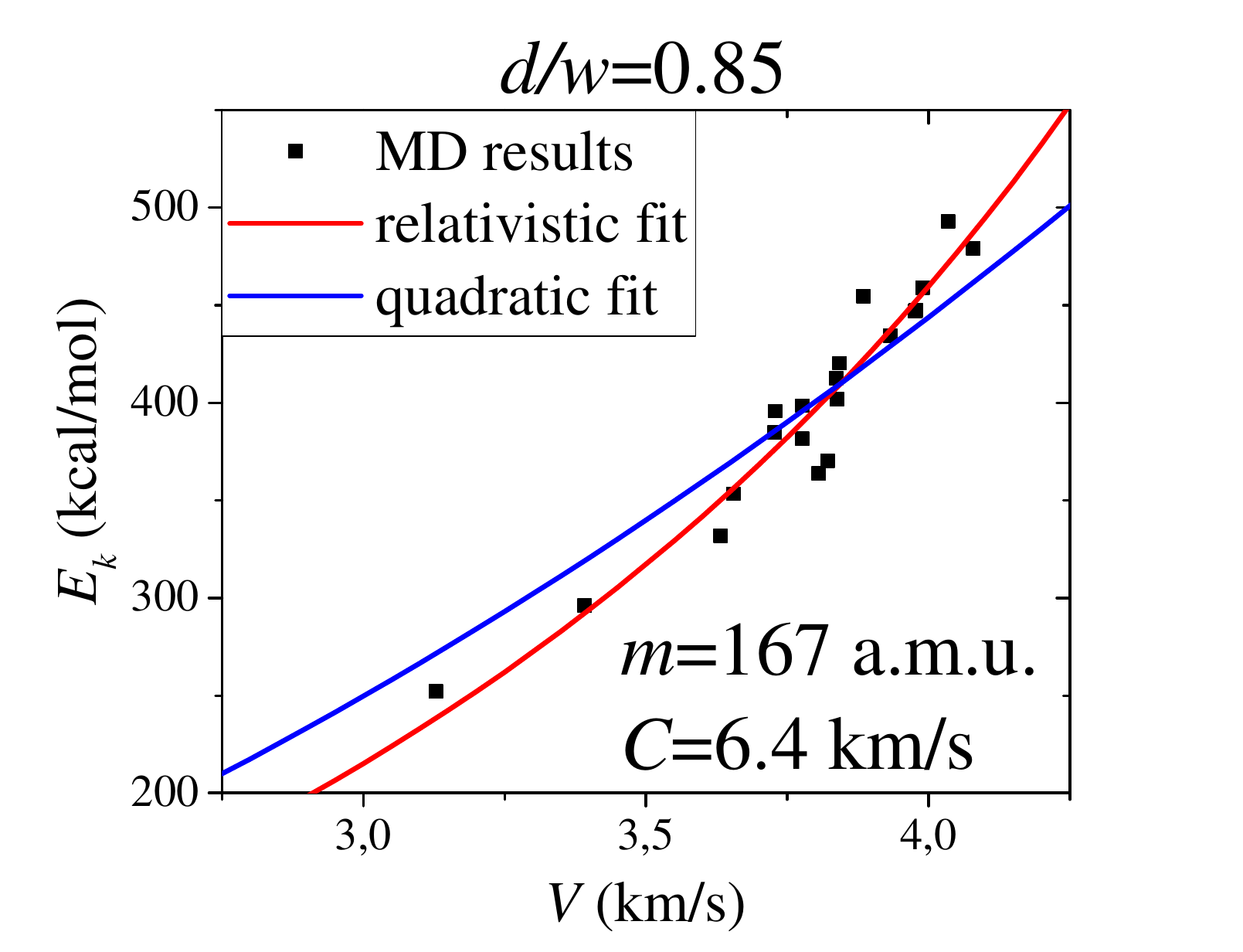}
\includegraphics*[width=55mm]{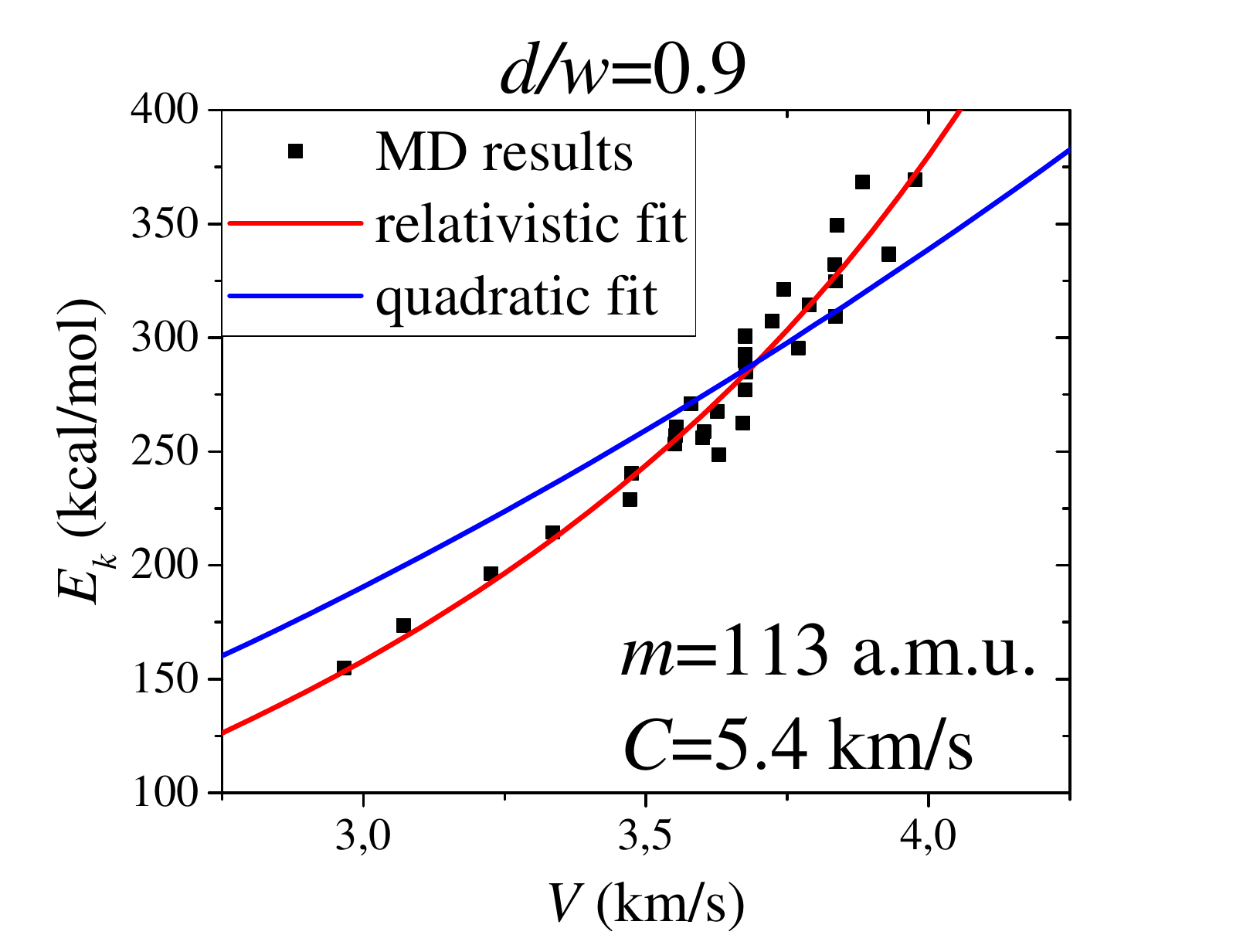}
\includegraphics*[width=55mm]{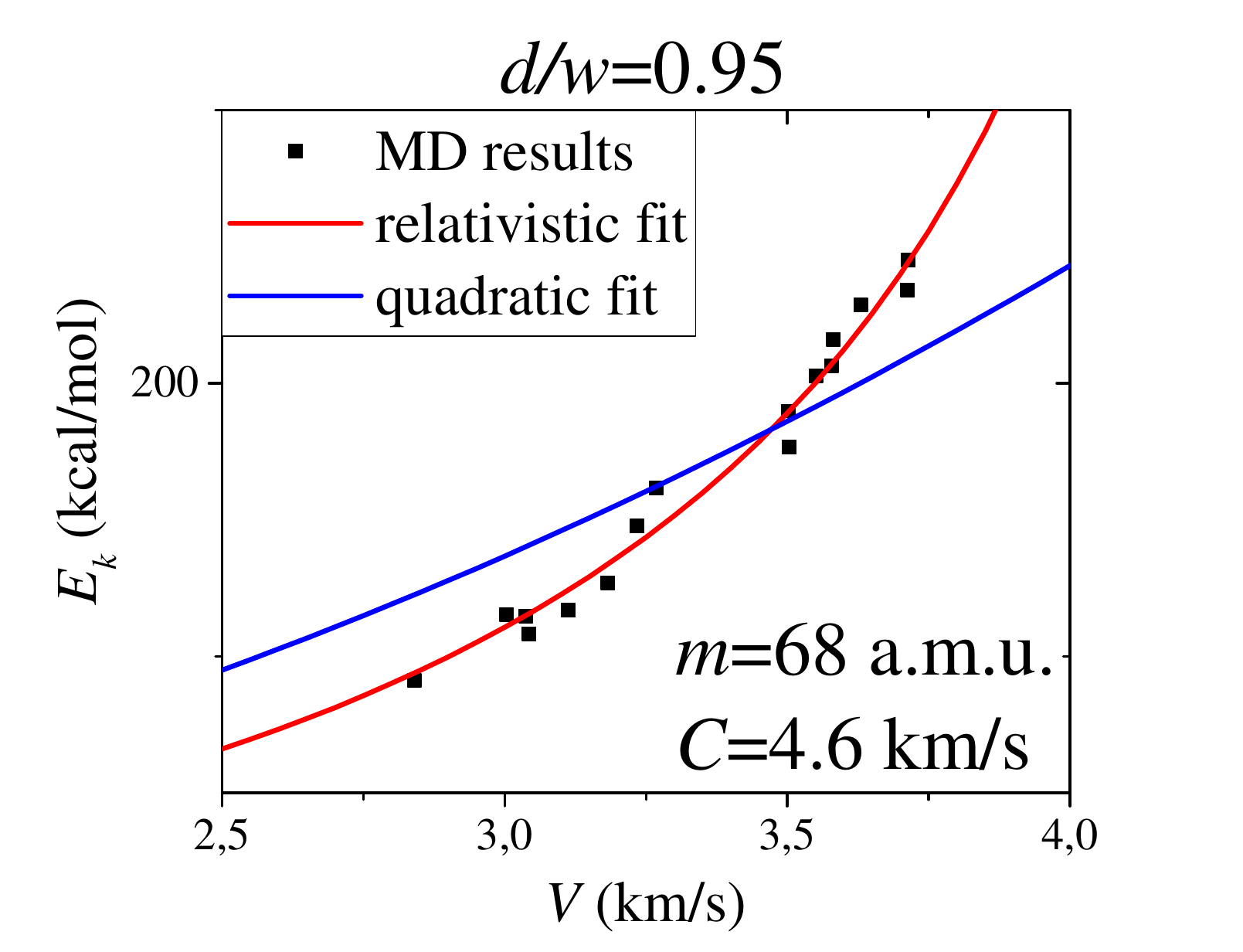}
\includegraphics*[width=55mm]{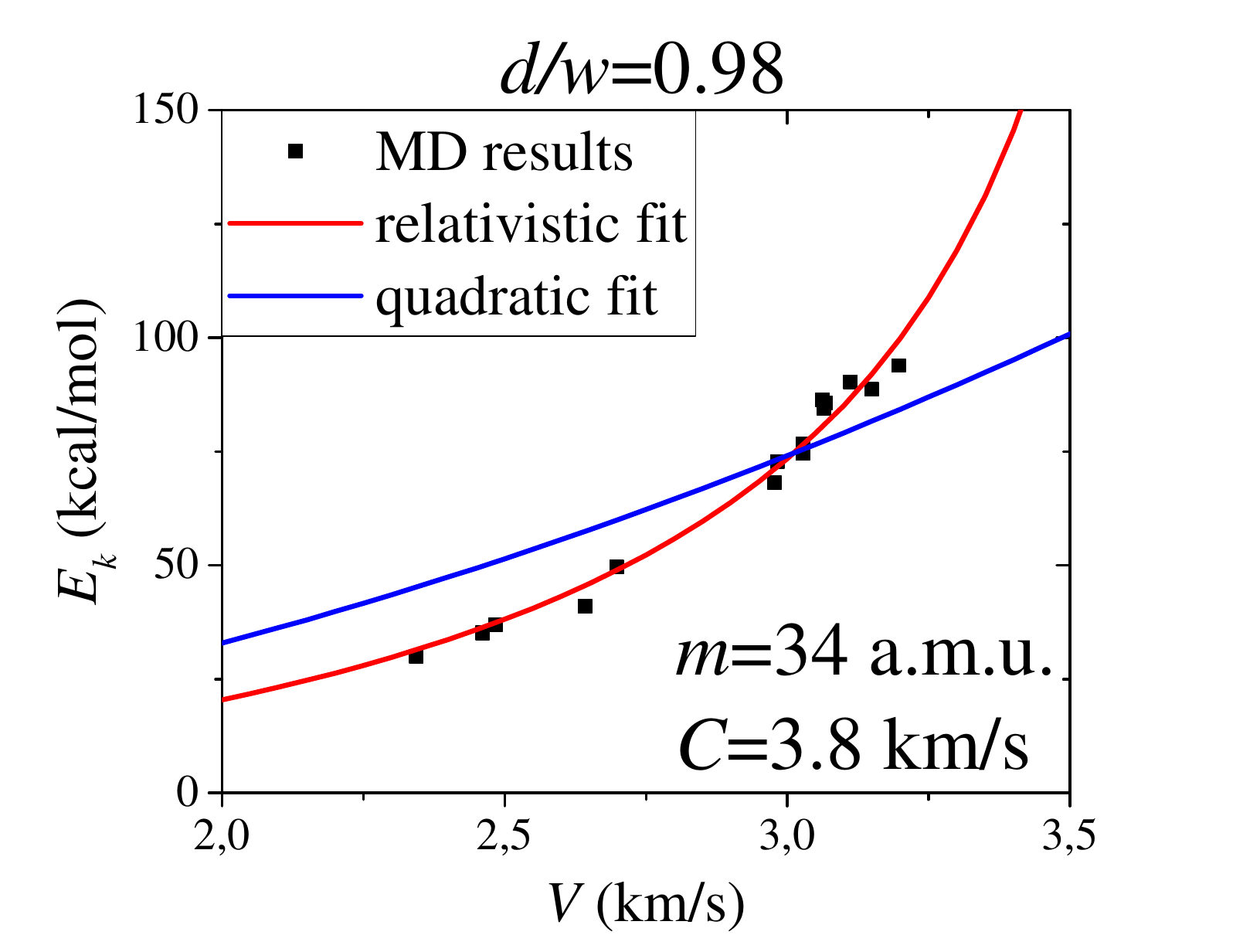}
\caption{Graphene kink kinetic energy in $w=9$ rings membrane at several values of $d/w$. The relativistic kinetic energy expression (red line) much better fits the MD data. The values of $m$ and $C$ are given for the relativistic fit (Eq.~(\ref{eq:Kink_Ek})).}
\label{fig:Ek}
\end{figure*}

The moving kinks were generated using Method 2 from Sec.~\ref{sec:MD} (for their generation a downward force was applied to a group of atoms near the left short edge of buckled up membrane), and their kinetic energies and velocities were extracted from molecular dynamics simulations. We have observed that $(\alpha,-)$-kinks are created when the pulling force exceeds a  threshold value. Moreover,
the initial speed of graphene kinks depends on the pulling force (saturating at about $~5$ km/s in $w=9$ rings, $d/w=0.9$ membrane~\cite{Yamaletdinov17b}) and stays practically unchanged when the kink moves along the membrane.
 Unfortunately, Method 2 generates a significant amount of noise that contributes to the total kinetic energy of  membrane. This makes challenging the precise measurement of the kink kinetic energy.
To reduce the noise, after $5$~ps of the initial dynamics, the atomic coordinates and velocities in the regions beyond the moving kink
(starting at $\pm20$~\r{A} from the kink center) were set to the values in the optimized buckled up or down membrane, and the pulling force was removed. Fig.~\ref{fig:Ek} shows the numerically found kinetic energy of kink as a function of its velocity fitted with the classical ($E_k=mV^2/2$) and relativistic (Eq.~\ref{eq:Kink_Ek}) expressions.

It is  interesting that both the kink mass $m$ and characteristic speed $C$  strongly correlate with each other: they both increase as the buckling increases. The numerical values of these parameters are  reasonable. While the characteristic speed $C$ is of the order of the speed of sound in flat graphene ($9.2-18.4$ km/s~\cite{Adamyan11a}), the effective kink mass $m$ is comparable to the mass of several atoms (one carbon atom mass equals $12.0107$ a. m. u.).

\subsection{Kink-antikink scattering} \label{sec:3c}

In Ref.~\cite{Yamaletdinov17b} we studied the kink-antikink scattering in a $w=9$ rings graphene membrane at $d/w=0.9$.
Using Method 2 (see Sec.~\ref{sec:MD} and Ref.~\cite{Yamaletdinov17b}), kinks and antikinks with the speeds in the interval  from $3$~km/s to $5$~km/s were created at the opposite sides of membrane and collided at the center. The results were compared with
those for the classical $\phi^4$ model in which, depending on the initial kink velocity, the kink-antikink collision leads either to the reflection, or annihilation with formation of a long-radiating bound state, or reflection through two- or several-bounce resonance collisions (see Fig.~\ref{fig:2} and Refs.~\cite{campbell1983,Anninos91a,kevrekidis2019dynamical}).

\begin{figure*}[tb]%
 (a)\includegraphics*[width=55mm]{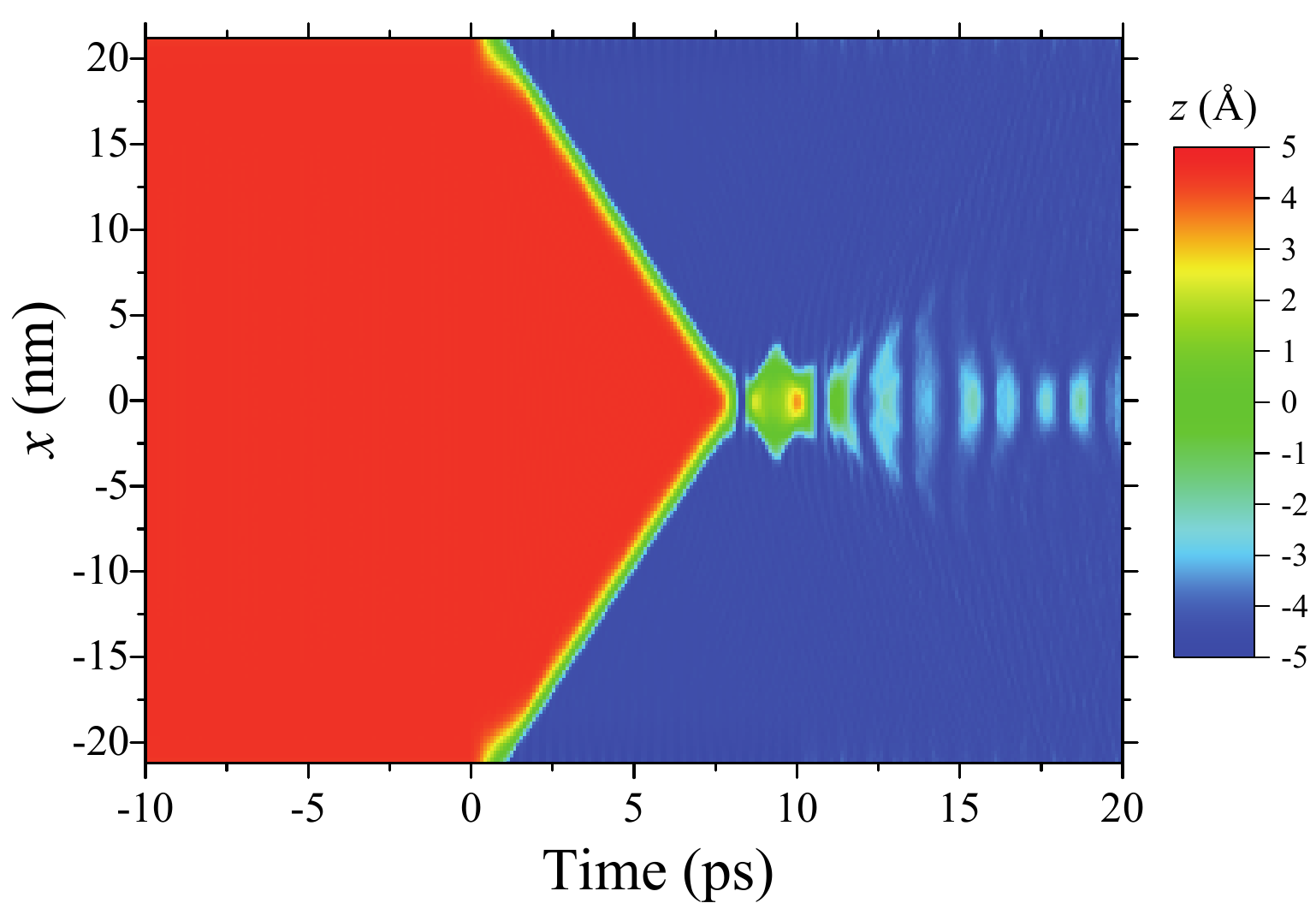}
 \;(b)\includegraphics*[width=55mm]{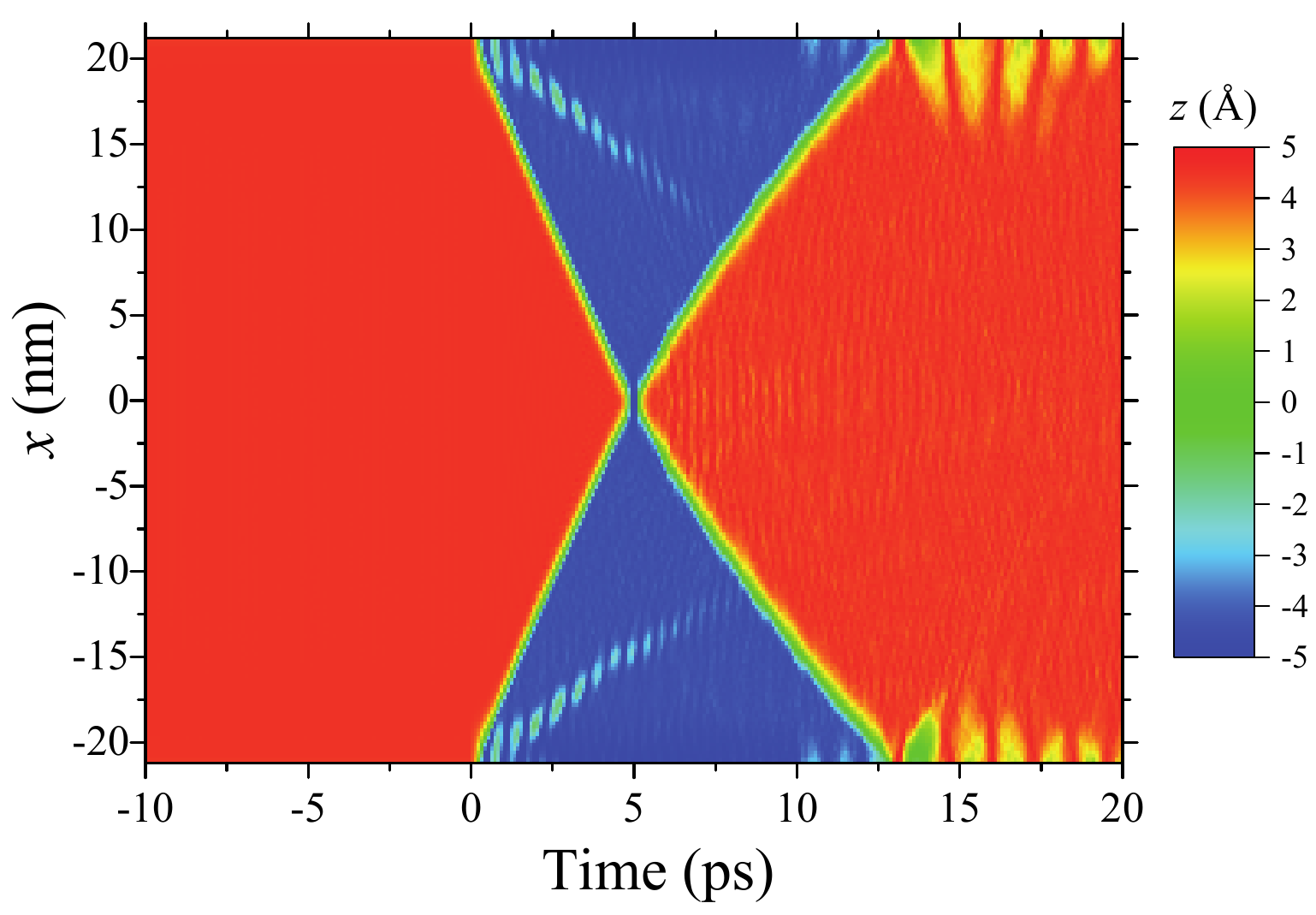}
\caption{ Collision of (a) slower and (b) faster moving kink and antikink. Reprinted with permission from Ref.~\cite{Yamaletdinov17b}.}
\label{fig:7}
\end{figure*}

\begin{figure}[tb]%
 \includegraphics[width=55mm]{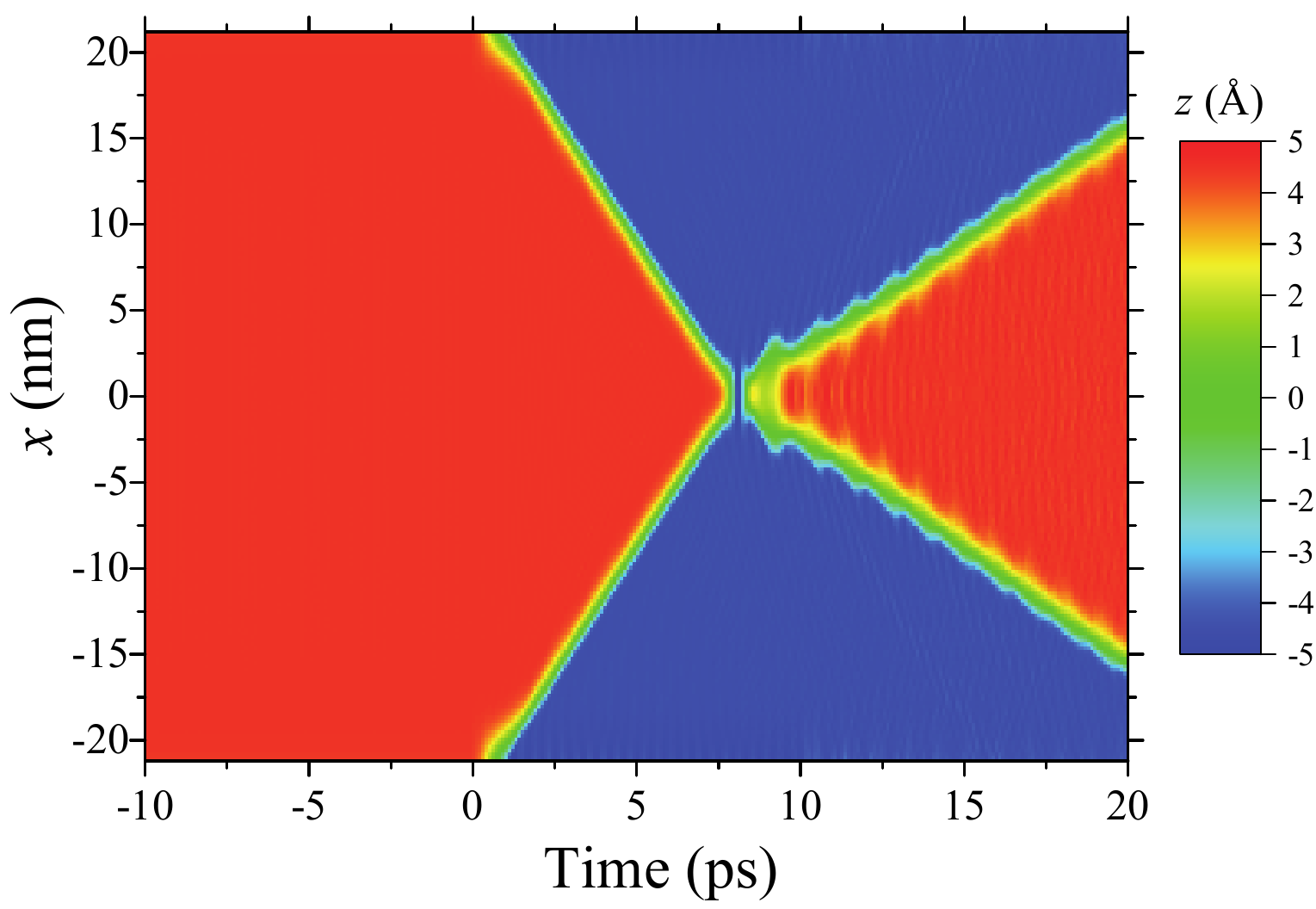}
\caption{ A resonance phenomenon in the kink-antikink scattering. Reprinted with permission from Ref.~\cite{Yamaletdinov17b}.}
\label{fig:7p}
\end{figure}

To study the kink-antikink scattering, we have performed a series of calculations in which the force used to generate moving kinks and antikinks was varied. It was observed that the decaying bound state is formed in the collisions of slower moving kinks and antikinks. An example of such situation is presented in Fig.~\ref{fig:7}(a) where the collisions occurs at the speeds of $~2.9$ km/s. We  also observed that the collision of fast moving kink and antikink leads to their immediate reflection (see Fig.~\ref{fig:7}(b)).

In Ref.~\cite{Yamaletdinov17b} we did not observe, however, a series of resonances below the critical velocity separating the annihilation and reflection regimes (for $\phi^4$ model resonances, see Fig.~\ref{fig:2}). At the same time, the behavior similar to the two-bounce reflection was spotted close to the critical velocity, see Fig.~\ref{fig:7p}. However, in some preliminary simulations of membranes with smaller $d/w$ we spotted a reflection window below the critical velocity. Further work is required to evaluate and refine this result as the effect may be related to noise (which is a significant side-effect in this type of simulations). Generally, the deviation of the scattering in graphene compared to the one in $\phi^4$ model can be explained by a larger number of energy relaxation channels/degrees of freedom in graphene due to its two-dimensional structure.

\subsection{Radiation-kink interaction} \label{sec:3d}

Recently the radiation-kink interaction was investigated by us in Ref.~\cite{Yamaletdinov19a}. The interest in this topic has been motivated by an unexpected theoretical prediction of a \textit{negative radiation pressure effect} (NRP) in $\phi^4$ field model~\cite{Tomasz08a}. In the standard linearized scattering theory the radiation-kink interaction is described by a second-order term $\sim A^2|R^2|$, where $A$ is the radiation amplitude, and $R$ is the reflection coefficient.
A distinctive feature of $\phi^4$ model is the absence of the radiation-kink interaction up to forth order in $A$, so that the force $F$ experienced by the kink $F\sim \pm A^4$. The NRP effect corresponds to the minus sign (the force is directed towards the radiation source), and can be explained as follows. In $\phi^4$ field model, the radiation-kink interaction is nonlinear and may lead to  frequency doubling. As the doubled-frequency (transmitted) waves carry more momentum than the incident ones, the kink must accelerate toward the radiation source to compensate the momentum surplus.

In graphene, the NPR effect is quite complex as the role of radiation is played by phonons, which can be of several types~\cite{Nika_2012}. Moreover, the boundary scattering as well as phonon-kink interaction may lead to the conversion between different types of phonons  -- an effect, which is beyond the scope of this investigation. A qualitative description of the phonon-kink scattering can be obtained in terms of a multichannel scattering model~\cite{Yamaletdinov19a}, in which the force $F$  is written as
\begin{equation}
    F\sim \mathcal{P}_i +\sum_j \frac{k_j}{k_i}(|R_{ij}|^2-|T_{ij}|^2)\mathcal{P}_i,
    \label{F_multi}
\end{equation}
where $k_{i}$ is the wave number, and it is assumed that the incoming radiation is contained in the $i$-th channel. In Eq.~(\ref{F_multi}), the first term is responsible for the absorption of the $i$-th component of incoming momentum flux $\mathcal{P}_i$, while the second term describes the emission of the absorbed radiation (from channel $i$) into the reflected and transmitted modes in $j$-th channels (with the scattering probabilities $|R_{ij}|^2$ and $|T_{ij}|^2$, respectively). A favorable condition for NRP is when the scattering probabilities $|R_{ij}|^2$ are small, and one of $|T_{ij}|^2$ into a certain high-momentum channel is large.

The negative radiation pressure effect in buckled graphene  was studied using Method 3 in Sec.~\ref{sec:MD}. To generate phonons, a
sinusoidal force was applied to a group of atoms near the left short edge of membrane (the radiation source). The kink was initially placed at a distance of $200$ \r{A} from the radiation source, and its position as a function of time was recorded. Due to the large space of parameters, we performed a series of MD simulations for a single buckled membrane ($w=9$ rings, $d/w=0.9$) varying the amplitude, frequency and direction of sinusoidal force. For additional information, see Sec.~\ref{sec:MD} or Ref.~\cite{Yamaletdinov19a}. The direction of coordinate axes can be found in Fig.~\ref{fig:3}.

\begin{figure*}[h]%
    \includegraphics*[width=55mm]{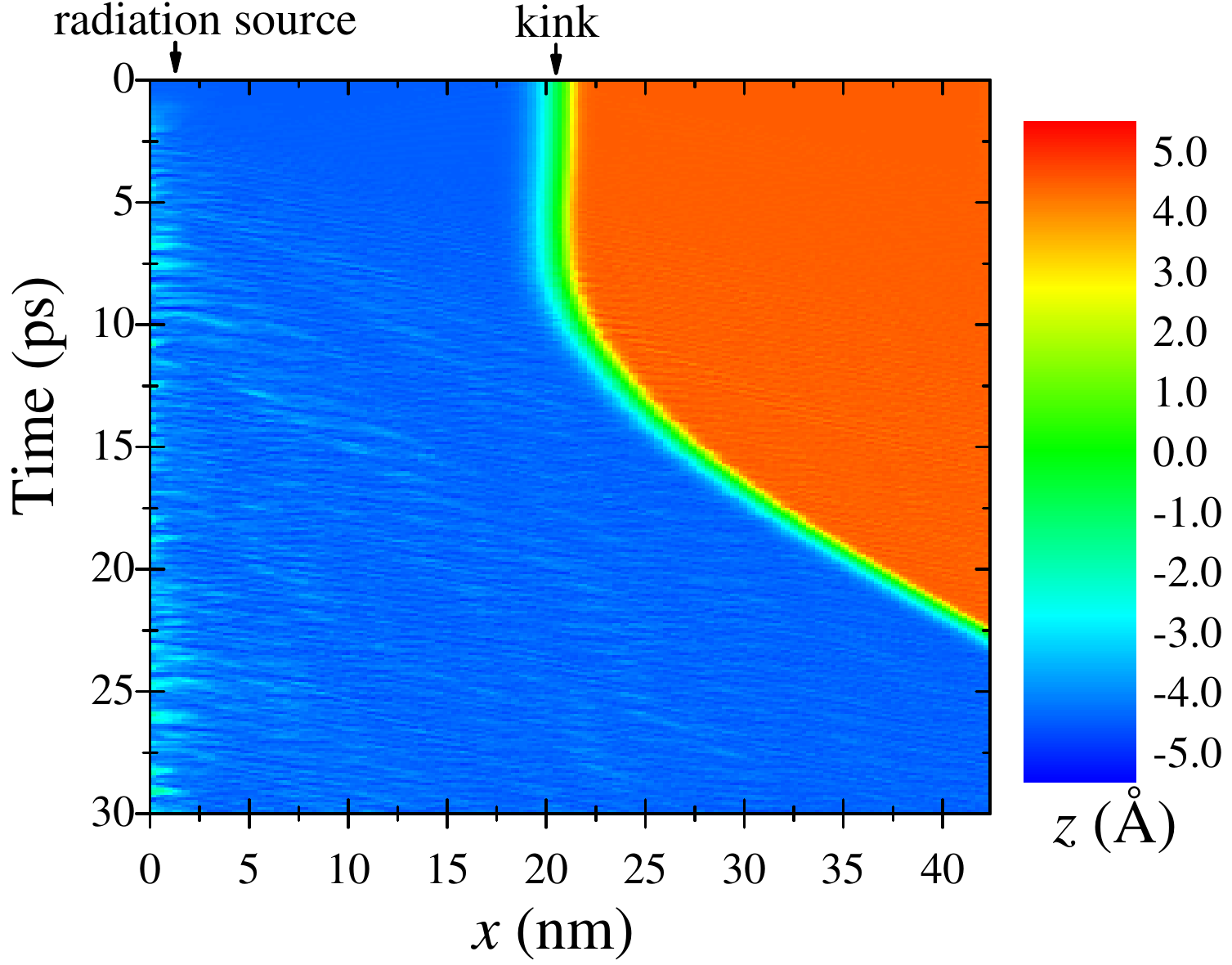}
    \includegraphics*[width=55mm]{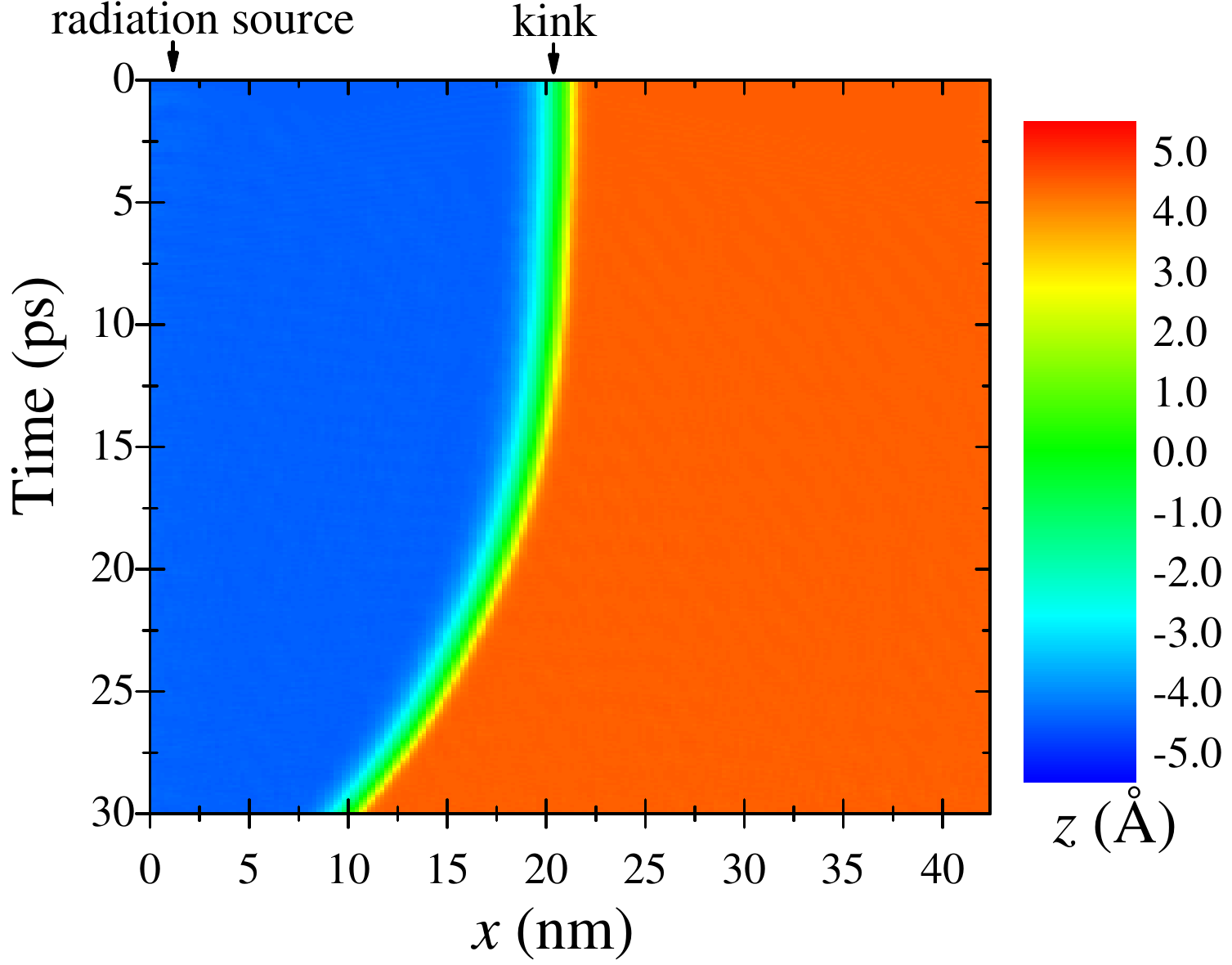}
    \caption{ Positive (left) and negative (right) radiation pressure effect. In both cases the radiation is caused by an external sinusoidal force in $y$-direction ($80$ pN/atom) with the period of $T=220$ fs (left) and $T=190$ fs (right).  Reprinted with permission from Ref.~\cite{Yamaletdinov19a}.}
    \label{fig:prp_nrp}
\end{figure*}

\begin{figure*}[h]%
    \includegraphics[width=115mm]{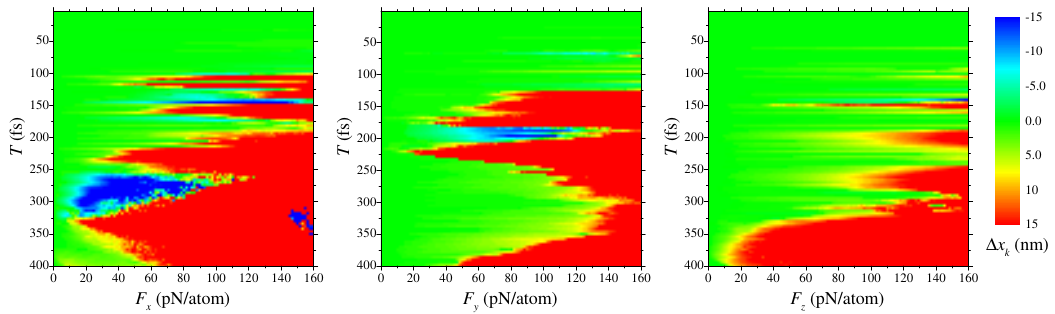}
    \includegraphics[width=115mm]{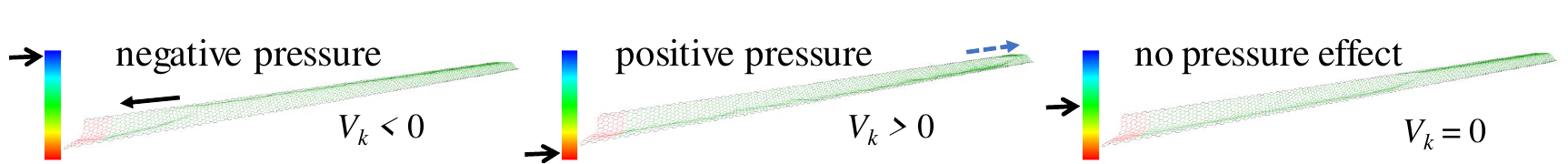}
    \caption{ The final kink displacement as a function of the applied force amplitude and period. The blue regions correspond to the negative pressure effect, the red ones - to the positive radiation pressure effect. Reprinted with permission from Ref.~\cite{Yamaletdinov19a}.}
    \label{fig:scan_nrp}
\end{figure*}

The positive and negative radiation pressure effects (PRP and NRP) are exemplified in Fig.~\ref{fig:prp_nrp}. The difference between these effects is in the direction of kink displacement: in the case of PRP the radiation pushes the kink away from the radiation source (located in the vicinity of $x\sim0$ in Fig.\ref{fig:prp_nrp}), while in the case of NRP the radiation pulls the kink towards the radiation source. The $\phi^4$ model predicts a very narrow frequency window for NRP~\cite{Tomasz08a}. To understand how the radiation pressure depends on the driving force parameters,
an extensive scan in the force amplitude-force period space for forces in the $x$-, $y$- and $z$-directions was performed. The results are shown in Fig.~\ref{fig:scan_nrp}. Here, the blue regions correspond to the NRP effect (except of $F_x\sim 150$~pN/atom $T\sim 350$~fs region in the left plot), while the red regions - to the PRP effect. We emphasize that the type of effect has a complex dependence on the driving force parameters, and is different for $F_x$, $F_y$ and $F_z$ excitations because of the different types of phonons produced by forces in different directions.

The negative radiation pressure effect that has been  observed in our MD results can be explained as follows. Assume that the incoming radiation has the wavelength $\lambda_{in}$. The non-linear interaction with kink scatters the incoming phonons into the modes with the wavelengths $\lambda_{tr,1}=\lambda_{in}$, $\lambda_{tr,2}=\lambda_{in}/2$, etc. The most efficient higher-harmonic generation can be expected when the incoming wave is in  resonance with the kink length $L_K$, namely, $L_K=n\lambda_{in}$ where $n$ is an integer. In such situation, the role of the emission at $\lambda_{tr,2}$  is increased, and the kink is pushed in the negative $x$-direction.

\begin{figure}%
    \includegraphics[width=0.7\textwidth]{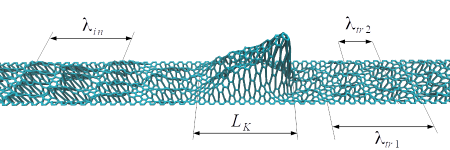}
    \caption{The scaled differences of atomic positions in the NRP regime induced by an $F_z=120$ pN/atom, $T=144$ fs force. Reprinted with permission from Ref.~\cite{Yamaletdinov19a}. }
    \label{fig:kink_nrp}
\end{figure}

To test the above speculation, it is convenient to consider the scaled difference of atomic positions. For the case of NPR effect, this quantity is plotted in Fig.~\ref{fig:kink_nrp} for $F_z=120$ pN/atom and $T=144$ fs. Fig.~\ref{fig:kink_nrp} clearly demonstrates that the above discussed conditions for NRP effect ($L\approx\lambda_{in}=42.6$ \r{A}, $\lambda_{tr,2}=\lambda_{in}/2$) are met in this specific MD simulation. Similar results were obtained for the driving force in $y$-direction ($F_y=74$ pN/atom and $T=194$ fs). It is important to note that in reality the radiation-kink scattering involves several channels, and single polarized monochromatic incoming waves may be partially transformed to waves with another polarization or/and with higher harmonics. In Ref.~\cite{Yamaletdinov19a} we also performed an analysis of vibrational modes of the kink and refer the interested reader to this publication for additional information.

\section{Kinks in longitudinally compressed graphene} \label{sec:4}

It is of interest to understand the effect of a longitudinal compression superimposed on the compression in the transverse direction.
This is not an abstract question because the longitudinal compression can be introduced during the fabrication stage in an experimental setup.
For instance, the buckled graphene can be created by leveraging
an intriguing property of graphene known as negative thermal expansion (opposite
to most materials, graphene contracts on heating and expands on cooling)~\cite{yoon2011negative}.
A thermally oxidized silicon wafer with an array of lithographically defined U-shaped grooves can be used as a substrate
(the grooves can be formed by chemical or plasma etching). Graphene transferred to the wafer surface at a
high temperature will cool, expand in {\it two} directions, and buckle above the grooves.

\begin{figure}[h]%
(a) \hspace{0.5cm}
 \includegraphics*[width=73mm]{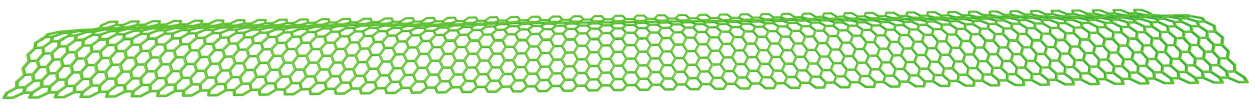} \hspace{0.5cm} $l/L=0.98$ \\
(b)
\hspace{0.6cm}
 \includegraphics*[width=71mm]{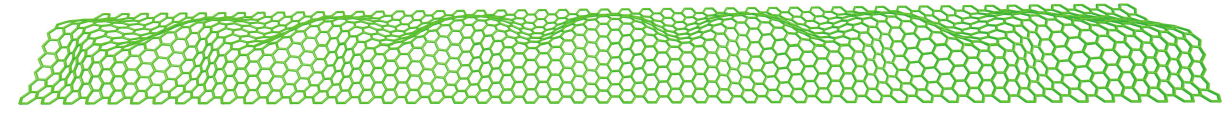} \hspace{0.6cm} $l/L=0.95$\\
(c)
\hspace{0.7cm}
 \includegraphics*[width=68mm]{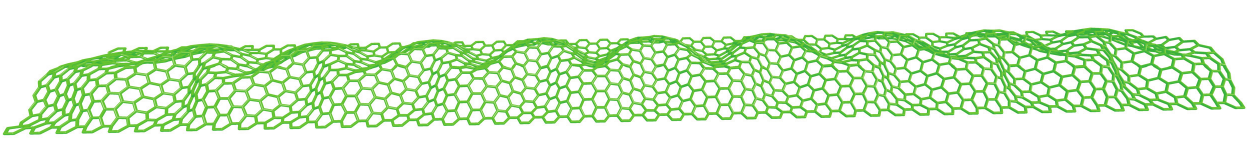} \hspace{0.8cm} $l/L=0.90$\\
(d)
\hspace{1.1cm}
 \includegraphics*[width=61mm]{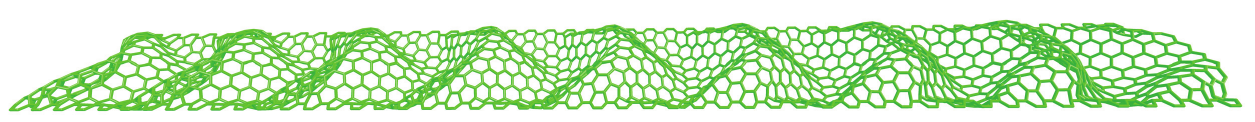} \hspace{1.1cm}$l/L=0.80$ \\
\caption{Low energy conformations of a buckled membrane at several degrees of longitudinal compression ($d/w=0.9$, $w=9$ rings).}
\label{fig:13}
\end{figure}

Molecular dynamics simulations of longitudinally compressed membranes were performed similarly to the simulations described above with the only difference that the initial longitudinal separation between the atoms were scaled by a factor of $l/L$, where $l$ is the length of compressed membrane. Some initial results on the properties of longitudinal compressed membranes were obtained using Methods 1 and 2 from Sec.~\ref{sec:MD}. Fig.~\ref{fig:13} shows examples of optimized geometries found with the help of Method 1 for several selected values of $l/L$ and fixed $d/w=0.9$.

\begin{figure}[h]%
\includegraphics[width=65mm]{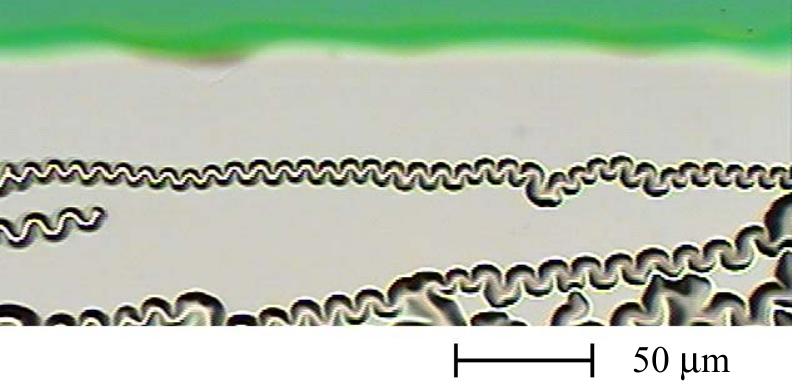}
\caption{Telephone cord buckling of a biaxially compressed tungsten film grown on silicon oxide.}
\label{fig:14}
\end{figure}

When the longitudinal compression is relatively small (below a critical value), the membrane is not visibly deformed as we show in Fig.~\ref{fig:13}(a). However, above the critical value (which is in the interval $\{0.95,0.98 \}$ of $l/L$ for our membrane), an instability develops strongly resembling the telephone cord buckling~\cite{moon2002,ni2017shape} in appearance (Figs.~\ref{fig:13}(b) and (c)). In the past, the telephone buckling instability has been observed in the delamination of biaxially compressed thin films~\cite{moon2002,ni2017shape} and have been extensively studied both experimentally and theoretically. Fig.~\ref{fig:14} exemplifies the telephone cord delamination using an in-house grown structure. The striking similarity of Fig.~\ref{fig:14} with Figs.~\ref{fig:13}(b) and (c) indicates that the deformation mechanism in the graphene membrane is likely of the same origin. When the longitudinal compression is very strong, the structure experience the transition to a different shape as shown in Fig.~\ref{fig:13}(d).

\begin{figure}[h]%
\includegraphics[width=55mm]{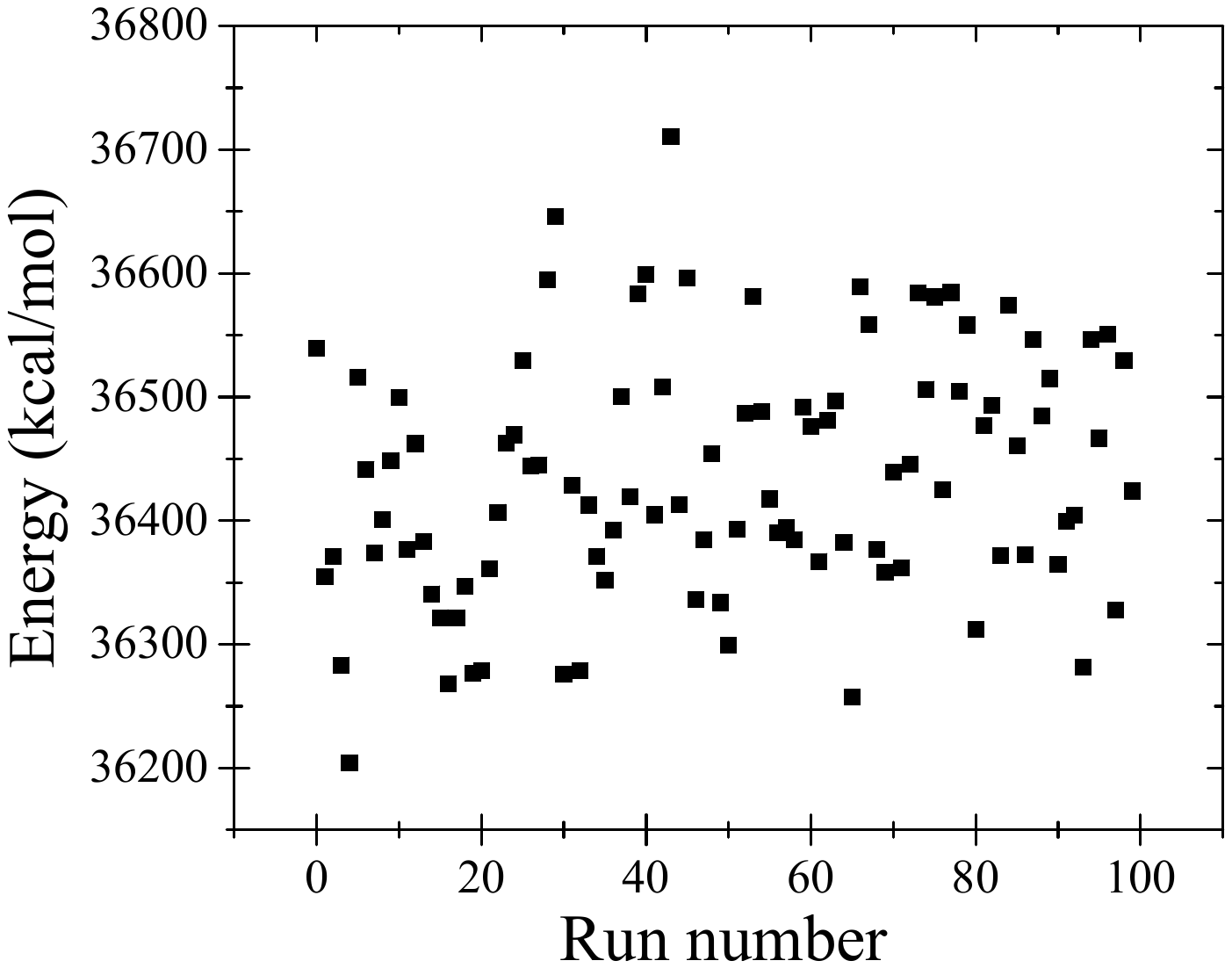}
\caption{Final configuration energy found in 100 runs of Method 1 simulation ($w=9$ rings, $d/w=0.9$, $l/L=0.8$).}
\label{fig:15}
\end{figure}

The longitudinal instability leads to significant variations in the stable conformation energies calculated using Method 1.
Fig.~\ref{fig:15}, for instance, demonstrates the absence of distinct energy states in $l/L=0.8$ membrane in distinction to the longitudinally uncompressed membranes,
 see Fig.~\ref{fig:5}. While possible explanations include the position-dependence of kink energy in longitudinally deformed membranes, and trapping in local minima within molecular dynamics optimization process,
 such answers at this time remain purely speculative.  We hope to clarify these in the future.

\begin{figure*}[t]%
 \includegraphics*[width=55mm]{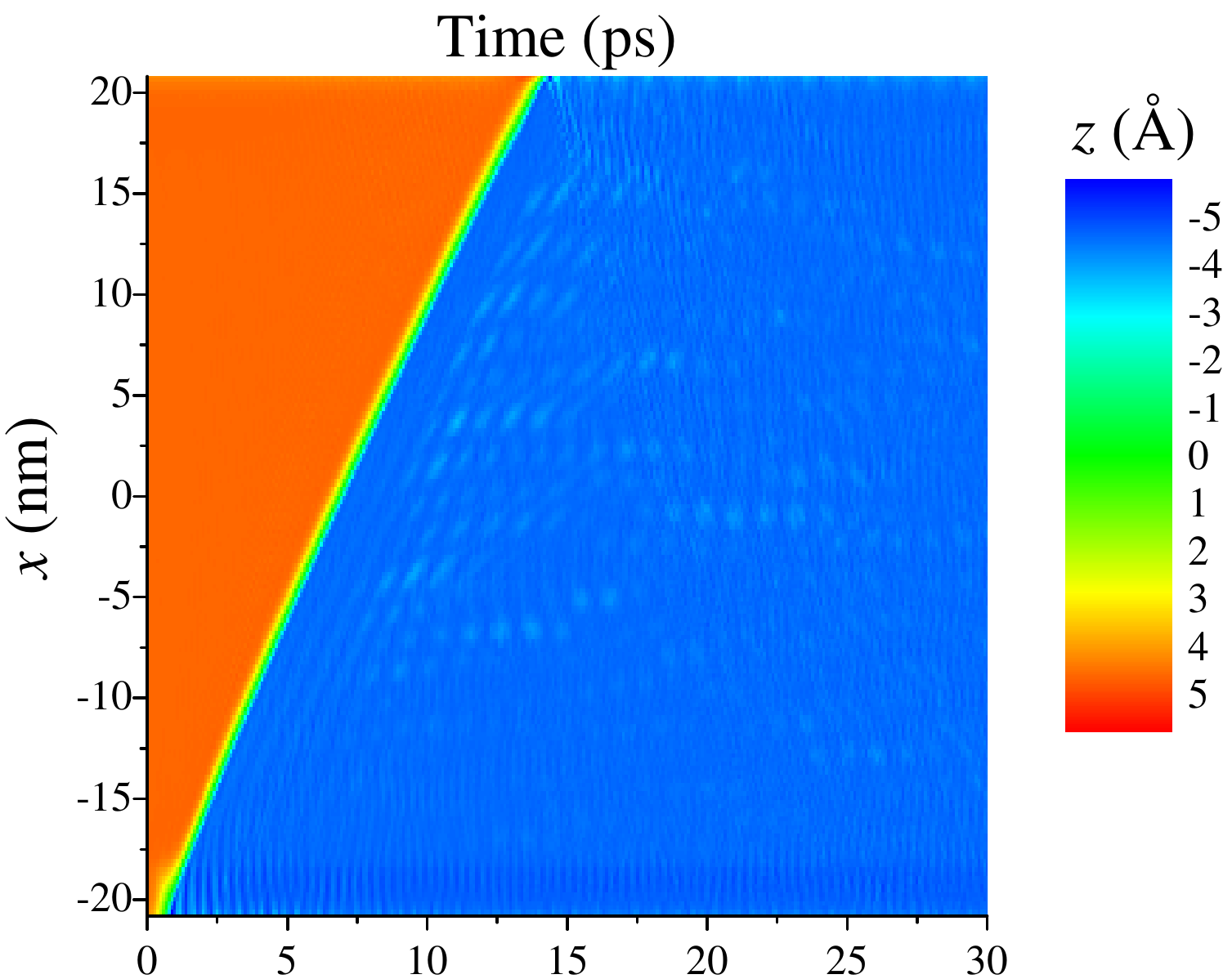}
 \includegraphics*[width=55mm]{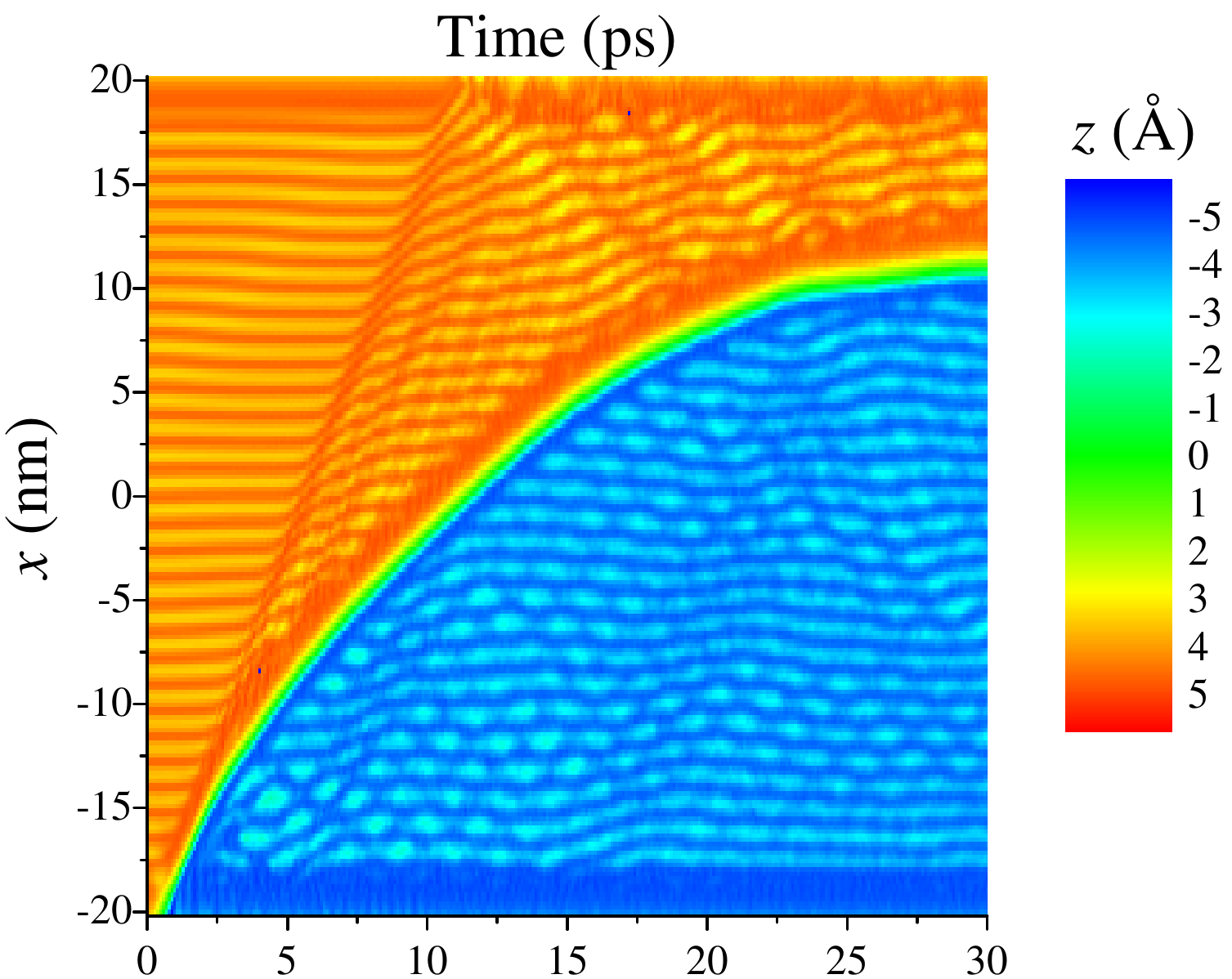}
\caption{Moving kinks in the longitudinally compressed membranes: (a) $l/L=0.98$, $d/w=0.9$, and (b) $l/L=0.95$, $d/w=0.9$. The kinks are created by $50$~pN/atom force.
}
\label{fig:16}
\end{figure*}

Method 2 was used to generate and study moving kinks in the longitudinally compressed membranes. It was observed that at small levels of compression (below the critical value)  kinks move at visibly constant speeds, see Fig.~\ref{fig:16}(a). However, kinks experience significant friction when the compression strength is above the threshold. An example of this situation is shown in Fig.~\ref{fig:16}(b), where the
kink velocity decreases almost to zero due to the friction effect. Moreover, a close examination of this plot reveals that  the kink pushes the telephone cord deformation as a whole in front of itself.

\begin{figure}[h]%
\includegraphics[width=55mm]{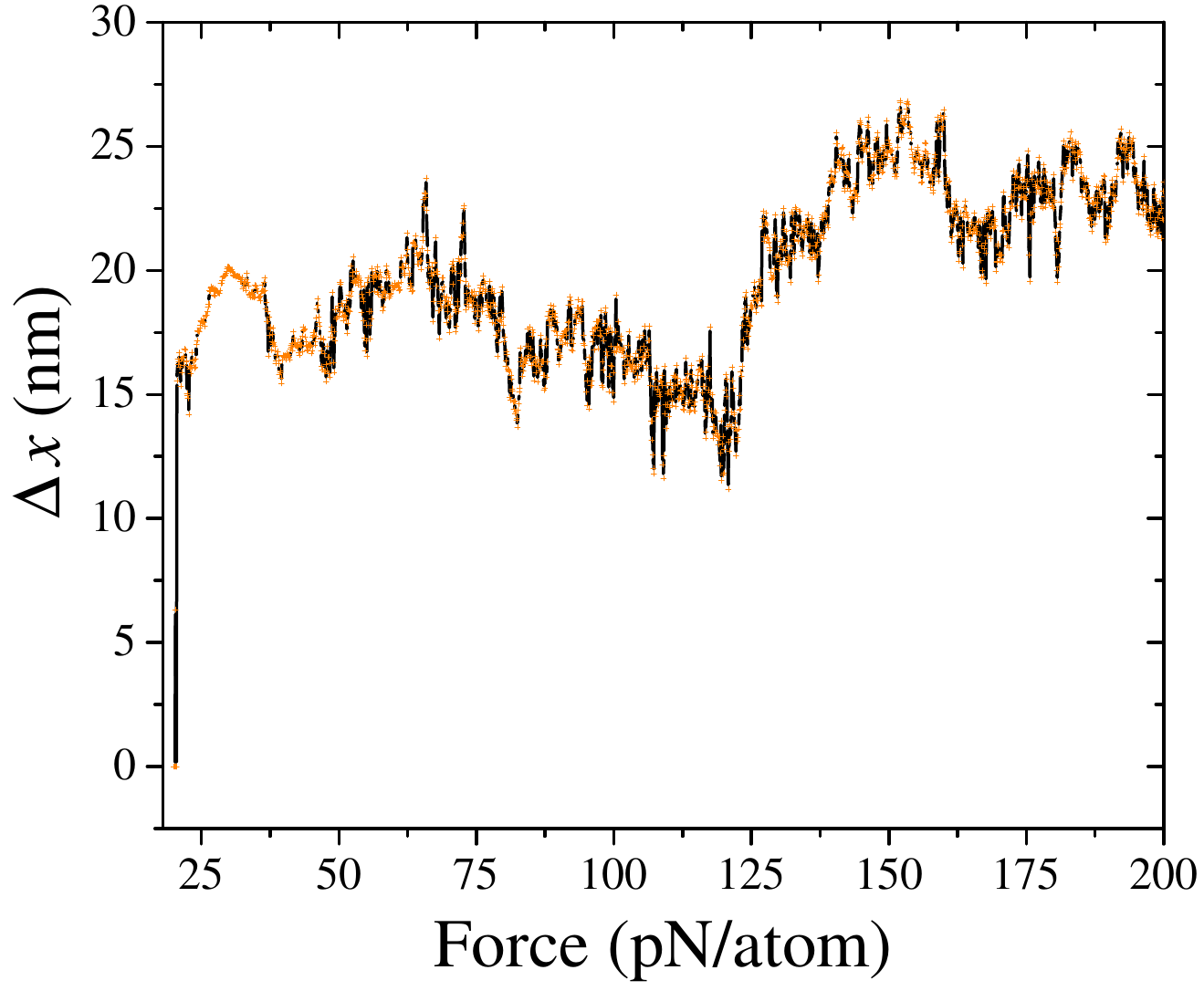}
\caption{Kink displacement within the simulation time interval (30 ps) as a function of the pulling force.}
\label{fig:18}
\end{figure}

To better understand the friction effect caused by the longitudinal buckling, we have performed a series of kink dynamics experiments using  five times longer membrane to avoid the interference of the reflected wave (from the right boundary) with the motion of kink within the simulation time. The kink position as a function of the pulling force was found using Method 2. One of the main observations is a considerable
 stochastic component of unclear origin in the kink dynamics. This observation is clearly manifested in Fig.~\ref{fig:18} that exhibits the final kink displacement as a function of pulling force.
 It should be emphasized that in each individual simulation the kink displacement is represented by a smooth curve. However, a small change in the pulling force sometimes leads to significant
 changes in the dynamical behavior. Moreover, we have attempted to fit the kink dynamics by a classical relativistic model with a velocity-dependent friction force of the form
 \begin{equation}\label{eq:friction}
   F_{d}=-a_{0 }\textnormal{sgn}(V)-a_1V-a_2V^2,
 \end{equation}
 where $a_i$ are fitting parameters, and $\textnormal{sgn}(\cdot)$ is the sign function. While for each individual trajectory we have obtained a very nice fitting, the fitting model parameters were inconsistent between different runs implying that the deterministic models are too narrow to account for the details of kink behavior in longitudinally buckled systems.

\section{Conclusion and outlook} \label{sec:5}

In this Chapter we have reviewed the properties of graphene kinks, and have extended the previous results in two directions.
First, we have explored the energetics of graphene kinks as a function of membrane width and degree of buckling. Second, we have studied the effect of  longitudinal compression on kink dynamics. Overall, our recent studies have uncovered a rich physics of topological excitations in buckled graphene membranes, including certain similarity with $\phi^4$ kinks, as well as some differences related to two-dimensional nature of graphene. We speculate that the graphene kinks may find applications in nanoscale motion because of their unique characteristics such as topological stability, high propagation speed, etc.

There are several future research directions to our work. First, it would be interesting to pursue the matter of
the scattering resonances  further, perhaps by more sophisticated calculations in which kinks and antikinks
 are generated with less noise. In particular, an important question is how the scattering depends on the types of kinks and antikinks involved and their symmetries. Second, an analytical 2D model of (at least) stationary kinks is highly desirable.
Such model could help understand the dependence of kink parameters (the energy and shape) on the geometrical and material properties of membrane.

Last but not least, the effects that we simulated on the nanoscale should be  also observed on the microscale (possibly in a modified form, e.g., with friction playing more important role) with quasi–two-dimensional (effective zero-thickness) materials.
Possible candidates include multilayer graphene, BN, MoS$_2$, copper oxides~\cite{Yin2016}, etc.  Therefore, the predictions that we have made
could be verified experimentally immediately.

\section*{Acknowledgments} The authors are thankful to T.~Roma\'nczukiewicz and V.~A.~Slipko for their
contribution to some of original publications reviewed here. The authors wish to thank  J. Kim and T. Datta
for their help with obtaining Fig.~\ref{fig:14}. R.~D.~Yamaletdinov gratefully acknowledges funding from RFBR (grant number 19-32-60012).


\newpage
\bibliographystyle{ieeetr}
\bibliography{memcap}



\Backmatter

\end{document}